\documentclass[twocolumn]{kj-article}

\usepackage[normalem]{ulem}

\let\oldsqrt\sqrt
\def\sqrt[#1]{\oldsqrt[\leftroot{3}\uproot{3}#1]}

\hypersetup{pdfauthor={Khaked Janada},pdftitle={Angular Control Charts}}

\title{Angular Control Charts\\\LARGE{A New Perspective for Monitoring Reliability of Multi-State Systems}}

\runtitle{Angular Control Charts: A New Perspective for Monitoring Reliability of Multi-State Systems}

\date{}

\author[a,*]{Khaled Janada\orcidlink{0000-0002-6522-8482}}
\author[a]{Hassan Soltan\orcidlink{0000-0003-2985-5072}}
\author[a]{Mohamed-Sobeih Hussein}
\author[a]{Ahmad Abdel-Shafi}

\affil[a]{Production and Mechanical Design Engineering Department, Faculty of Engineering, Mansoura University, Mansoura 35516, Egypt}
\affil[*]{Corresponding author: Khaled Janada. Email: khaledjanada@mans.edu.eg}

\begin{document}

\abstract{
	Control charts, as had been used traditionally for quality monitoring, were applied alternatively to monitor systems' reliability. In other words, they can be applied to detect changes in the failure behavior of systems. Such purpose imposed modifying traditional control charts in addition to developing charts that are more compatible with reliability monitoring. The latter developed category is known as probability limits control charts. The existing reliability monitoring control charts were only dedicated to binary-state systems, and they can't be used to monitor several states simultaneously. Therefore, this paper develops a design of control charts that accommodates multi-state systems, called here as the Angular Control Chart, which represents a new version of the probability limits control charts. This design is able to monitor state transitions simultaneously and individually in addition. Illustrative system examples are implemented to explore the monitoring procedure of the new design and to demonstrate its efficiency, effectiveness, and limitations.
}


\maketitle

\begingroup
\renewcommand{\arraystretch}{1.2}

\section*{Abbreviations}

\begin{supertabular*}{1\textwidth}{l@{}l@{ }p{0.37\textwidth}}
	ACC 		&:& Angular Control Chart \\
	ACL			&:& Angular Control Limit \\
	\sout{ACL}	&:& Angular Center Line \\
	ALCL 		&:& Angular Lower Control Limit \\
	AUCL 		&:& Angular Upper Control Limit \\
	CCC			&:& Cumulative Count of Conforming\\
	CQC 		&:& Cumulative Quantity Control \\
	MSS 		&:& Multi-State System \\
	MTTF 		&:& Mean Time-to-Failure \\
	TBE 		&:& Time-Between-Events \\
	TTF 		&:& Time-to-Failure \\
\end{supertabular*}

\section*{Nomenclature}
\begin{supertabular*}{1\textwidth}{l@{}l@{ }p{0.37\textwidth}}
	%\begin{tabularx}{1\textwidth}{l@{} l@{ } l}
	$c$ 				&:& Acceptable probability of false alarm \\
	S$_n$ 				&:& State transition from the $0^{th}$ state to the $n^{th}$ state \\
	$F^{-1}(p)$ 		&:& Quantile function of $p$ \\
	$T_C$				&:& Quantile at $p = 1/2$ (center line of original $t$-chart) \\
	$T_L$ 				&:& Quantile at $p = c/2$ (lower control limit of original $t$-chart) \\
	$T_U$ 				&:& Quantile at $p = 1 - c/2$ (upper control limit of original $t$-chart) \\
	$\alpha$ 			&:& Scale parameter \\
	$\beta$ 			&:& Shape parameter \\
	$\rho(a,b)$ 		&:& Ratio of quantiles function of $a$ and $b$ \\
	$\Phi^{-1}(p)$ 		&:& Standard normal quantile function of $p$ \\
	$\theta_C$ 			&:& Angle of \sout{ACL} \\
	$\theta_L$ 			&:& Angle of ALCL \\
	$\theta_U$ 			&:& Angle of AUCL \\
	%\end{tabularx}
\end{supertabular*}
\endgroup

\section{Introduction}
The statistical process control chart is a simple, yet effective, tool for monitoring, evaluating, and improving the performance of systems. The implementation of control charts generally involves identifying chance causes of variation (i.e., random causes), in a certain system behavior, and distinguishing them from assignable causes. The traditional Shewhart control charts were based on the normal approximation of the actual probability distribution of the system behavior. These charts use $k\dash\sigma$ control limits for monitoring variations in a process and detecting the occurrence of out-of-control status \citep{Sharma.2006, Besterfield.2013}. In addition to the traditional role of control charts in quality applications, they were further adapted for use in other fields.

Using control charts in reliability applications has gained growing attention in recent literature \citep{Ali.2016, Soltan.2019, SabriLaghaie.2022}.

If traditional Shewhart control charts are applied with skewed data, several inconveniences arise, including are: an inherent high probability of false alarms, requiring a large sample size for meaningful results, meaningless control limits, and the inability to detect changes in the process \citep{Xie.2002.Book, Sharma.2006, Ali.2016}. Since the failure data are severely skewed, it cannot be precisely approximated by the normal distribution. For such data, the estimated lower control limit of Shewhart charts usually turns out to be negative, and therefore is set to zero. Having a non-positive lower control limit inhibits the ability of the control chart to detect process improvements \citep{Chan.2000}.

Several procedures were proposed to enable using Shewhart control charts in reliability monitoring based on normalizing transformations of non-normal data. Numerous studies have been dedicated to finding the best transformation for failure time data \citep{Nelson.1994, Chou.1998, Yang.2000, Shore.2001}. The suitability of the chosen transformation type depends on the original data. Commonly used transformations include logarithmic transformation, Box-Cox (power) transformation, square root transformation, etc. The transformed data are then plotted on a traditional Shewhart control chart. For instance, \citet{Batson.2006} proposed modified procedures for using the traditional $\overline{X}$ and $s$ charts with exponential, Weibull, and lognormal failure data through conversion to the normal distribution. These procedures aim to enhance the performance of the traditional Shewhart charts when monitoring non-normal data. However, the implementation of these transformations and modified procedures still depended on the traditional $k\dash\sigma$ control limits. Therefore, these techniques, especially for heavily skewed data with large shifts or linear change in the \emph{mean time-to-failure} (MTTF), can still exhibit poor performance.

\textbf{\emph{Time-Between-Events} (TBE) control charts} were developed to avoid the limitations of traditional control charts when monitoring heavily skewed data. There are two main categories of TBE control charts. The first category is the Memory-type TBE control charts. Control charts in this category rely on using a cumulative statistic, such as the exponentially weighted moving average (EWMA) statistic, the cumulative sum (CUSUM) statistic, or other variations. The second category uses probability control limits to monitor TBE. This category includes the cumulative quantity control (CQC) chart, $t$-chart, $t_r$-chart, and other variations. For more detailed discussions about the numerous types of TBE control charts, see \citet{Liu.2006}, \citet{Ali.2016}, and \citet{Soltan.2019}.

Generally, Memory-type TBE charts outperform the probability limits charts in cases of small and moderate process shifts, and/or when these shifts can be accurately predicted. On the other hand, probability limits charts are more effective when process shifts are large and/or unpredictable. Another advantage of probability limits charts is that they are easier to design and implement \citep{Ali.2016, Fang.2016}. Therefore, the probability limits TBE control charts are adopted here as a base because of their suitability to the current purpose.

\textbf{Probability limits control charts} was first developed by \citet{Calvin.1983} who established a cumulative count control chart, based on the geometric distribution, for ``zero defect'' processes. This idea was further developed as the CCC (Cumulative Count of Conforming) chart. The CCC chart observes the number of conforming units between two defects and uses probability control limits instead of the traditional $k\dash\sigma$ limits \citep{Xie.2002.Book}. \citet{Chan.2000} developed the CQC (Cumulative Quantity Control) chart for application in continuous production. In the CQC chart, the quantity inspected between two defects can be measured by length, area, weight, etc., contrary to the CCC chart where the quantity inspected has to be measured in discrete units. Commonly, the occurrence of a defect in continuous high-yield processes is modeled as a homogenous Poisson process. Therefore, the distribution of the number of inspected units between two successive defects can be approximated by the exponential distribution \citep{Sharma.2006}.

\begin{figure*}[!h]
	\centering
	\includegraphics[width=0.75\linewidth]{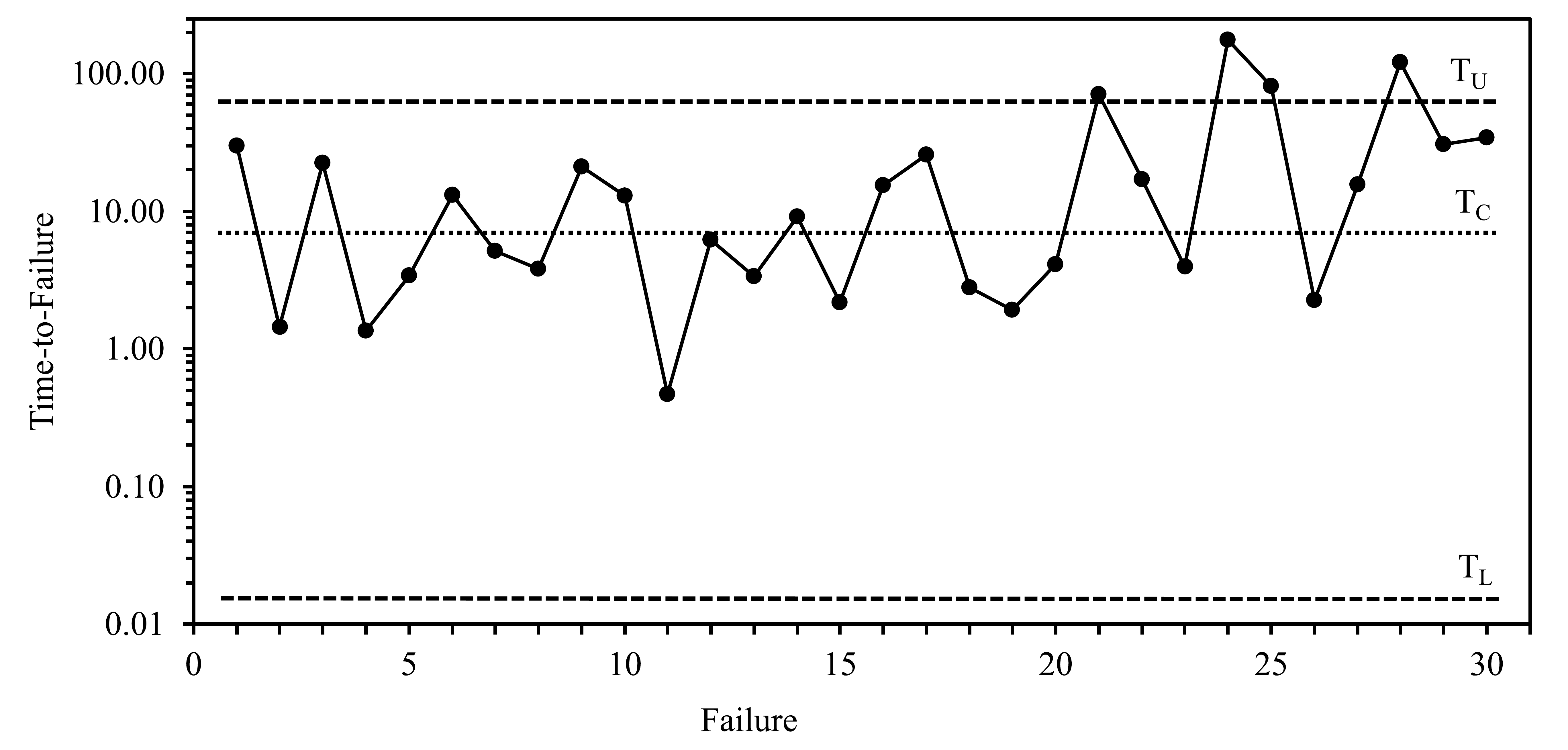}
	\caption{A typical $t$-chart \citep{Xie.2002.Article}}
	\label{Fig:tchart}
\end{figure*}

\citet{Xie.2002.Article} adapted the CQC chart for monitoring \emph{time-to-failure} (TTF), or in other words, the time between two successive failures, and named the adapted version the $t$-chart (\cref{Fig:tchart}). The center line, lower control limit, and upper control limit of the $t$-chart are set on the quantiles $T_C$, $T_L$, and $T_U$, respectively; those are defined as
\begin{align}
	T_C &= F^{-1} \paran{\frac{1}{2}} \label{Eq:TC}\\
	T_L &= F^{-1} \paran{\frac{c}{2}} \label{Eq:TL}\\
	T_U &= F^{-1} \paran{1-\frac{c}{2}} \label{Eq:TU}
\end{align}
where $c$ represents the acceptable probability of false alarm and $F^{-1}(p)$ is the quantile function (i.e., inverse cumulative distribution function).

\citet{Xie.2002.Article} also extended the $t$-chart to the $t_r$-chart that monitors the cumulative time until $r$ failures occur modeled by Erlang distribution (with a shape parameter $\beta = r$). The $t_r$-chart has higher sensitivity but unfortunately, it delays the detection of process degradation until $r$ occurrences of failure. \citet{Xie.2002.Article} originally applied the $t$-chart for monitoring failure data which were modeled with exponential and two-parameter Weibull distributions. Several adaptations and/or extensions of probability limits TBE control charts were introduced based on the $t$- and $t_r$-charts as those shown in \cref{Tab:Adaptations}.

\begin{table*}[!hbt]
	\caption{Adaptations and/or extensions of the $t$- and  $t_r$-charts}  \label{Tab:Adaptations}
	\begin{tabularx}{\textwidth}{
			>{\hsize=0.45\hsize\linewidth=\hsize}X
			>{\hsize=1.55\hsize\linewidth=\hsize}X}
		\toprule
		Source & Contribution \\
		\midrule

		\citet{Surucu.2009}&
		Extending the application of $t$- and  $t_r$-charts to the three-parameter Weibull distribution.\\

		\citet{Wu.2009} and \newline\citet{Wu.2009b}&
		Designing a TBE chart for monitoring both the time to failure $T$ and the magnitude of failure $X$. The chart is based on the event ratio statistic $R=X/T$.\\

		\citet{Farouk.2011}&
		Using the three-parameter Weibull $t_r$-chart for monitoring the behavior of systems with degraded reliability.\\

		\citet{Zhang.2011}&
		Introducing an economic design of exponential TBE control charts for monitoring multistage manufacturing processes.\\

		\citet{Xie.092011} and \newline\citet{Xie.2012}&
		Designing a new circle chart for monitoring periodic TBE measurements. In this chart, two concentric circles represent the lower and upper control limits, and the distance from the origin represents the observed measurement.\\

		\citet{Alsyouf.32015}&
		Using Triangular- and Weibull-based control charts for monitoring the reliability of repairable systems.\\

		\citet{Rasay.2020}&
		Applying the $t_r$-chart (based on the Erlang distribution) to monitor reliability tests, with replacement, that are censored at the $r^{th}$ failure. In this testing scheme, $n$ items are simultaneously tested, each with an exponential lifetime distribution. Once an item fails it is replaced by a new item, and the test is terminated after observing $r$ failures.\\

		\citet{Wu.2020}&
		Extending the application of $t$- and  $t_r$-charts to the three-parameter Fr\'echet distribution.\\

		\bottomrule
	\end{tabularx}
\end{table*}

In classical reliability theory, systems are viewed as having binary states—functioning or failed. This oversimplified view cannot be accurately applied to many complex real-life systems, which necessitates using more flexible and accurate reliability models. Consequently, the \emph{multi-state system} (MSS) reliability theory has been introduced \citep{Natvig.2010}. MSSs can exhibit several finite levels of performance ranging from perfect functioning to complete failure. The binary-state view of systems can be thought of as a special case of the MSS view in which the system has only two states \citep{Lisnianski.2010}.

Since the 1970s, numerous research studies have been devoted to the subject of MSS reliability, including reliability modeling, reliability assessment, reliability optimization, and maintenance planning of MSSs \citep{Lisnianski.2010, Yingkui.2012}. However, reliability monitoring approaches discussed earlier have not been extended to MSSs’ applications until the year 2015 when \citet{Genada.2015} introduced a new design of control charts for monitoring the failure behavior of MSSs. This design had modified probability control limits called \emph{Angular Control Limits} (ACLs) and was applied to MSSs with exponentially distributed state transitions.

This paper introduces control charts for reliability monitoring of MSSs with state transitions that can be modeled with any non-negative continuous probability distributions. This work builds on and further develops the design of control charts that was introduced by \citet{Genada.2015}. This new design of control charts, termed here as the \emph{Angular Control Chart} (ACC), can be used to monitor the overall reliability behavior of an MSS, as well as the reliability behavior of each of its states separately. An open-source software named \emph{AC Charts Software} was developed using C\# for easy implementation of ACCs. The code and software are available, under the MIT license, on GitHub (\url{https://github.com/Khaled-Janada/AC_Charts}) and archived in Zenodo \citep{Janada.2022}.

The remainder of this paper is organized as follows. \cref{Sec:StandardACC} demonstrates the standard design of the ACC. A generalized design of ACC is developed in \cref{Sec:GeneralACC}. \cref{Sec:Examples} exhibits the proposed ACC design through simulated examples. \cref{Sec:Summary} summarizes the configurations and salient findings of the standard and generalized designs of ACC. Concluding remarks are reported in \cref{Sec:Conclusions}.

\section{Standard Design of ACC} \label{Sec:StandardACC}
\begin{figure*}[!h]
	\centering
	\includegraphics[width=0.85\linewidth]{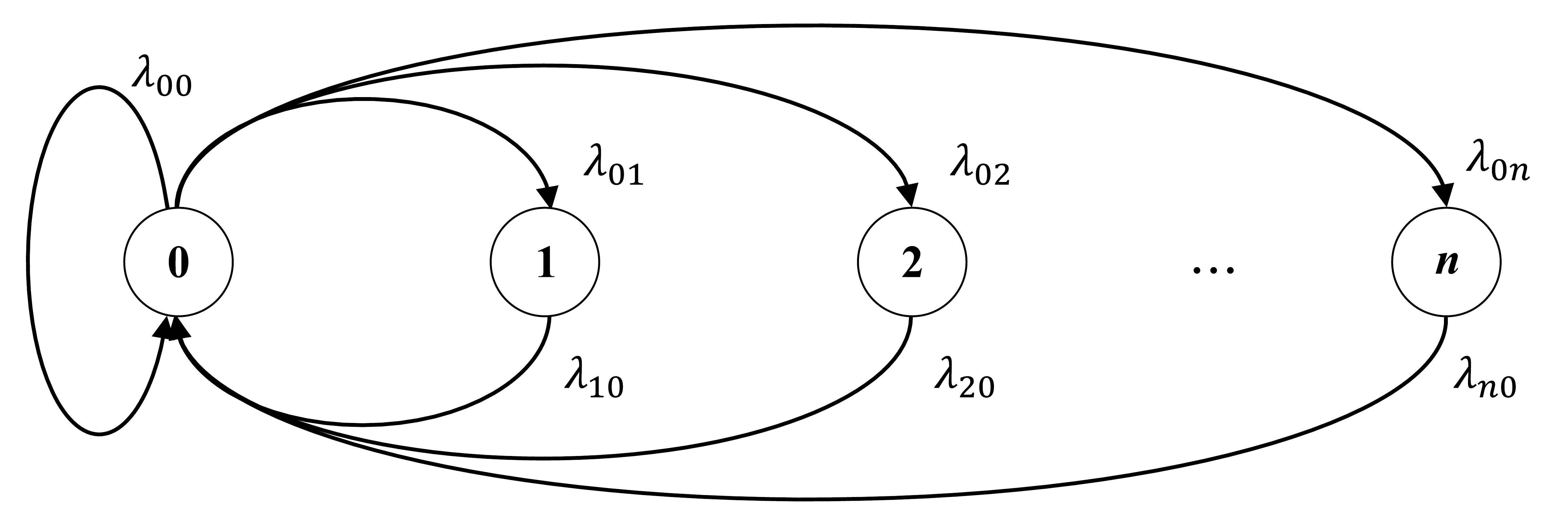}
	\caption{State transition diagram of a fully repairable MSS with only major failures}
	\label{Fig:MSS}
\end{figure*}

The system considered here is a fully repairable MSS having only major failures. Such a system has a perfect-functioning state (denoted as the $0^{th}$ state) in addition to $n$ states representing different levels of performance, such that all states are independent. In its $0^{th}$ state, the system can transition to any of the other states (from $1$ to $n$); these state transitions can be modeled with non-negative continuous probability distributions. Let S$_n$ denote the state transition from the $0^{th}$ to the $n^{th}$ state. The system is assumed fully repairable with no minor failures. This means that after each failure the system is immediately repaired (repair time is neglected) and restored to its $0^{th}$ state. There are no allowed state transitions from a failure state to another failure state with a lower level of performance (i.e., minor failures). The state transition diagram of this type of MSS is shown in \cref{Fig:MSS}, where $\lambda_{ij}$ represents the transition rate (failure rate) from state $i$ to state $j$.

The main problem with monitoring the reliability of an MSS, like the one shown in \cref{Fig:MSS}, is dealing with its varying transition rates (failure rates). The center line and probability control limits of the exponential $t$- and $t_r$-charts (\crefrange{Eq:TC}{Eq:TU}) depend on the failure rate (for the exponential distribution, the failure rate is the reciprocal of the distribution’s scale parameter). This means that different state transitions will have different control limits, and therefore, they cannot be monitored on the same $t$-chart. This is also true for non-exponential probability distributions. Thus, the original design of the $t$-chart lacks the means to adequately present and monitor the different states of an MSS.

\begin{figure*}[!h]
	\centering
	\includegraphics[width=0.8\linewidth]{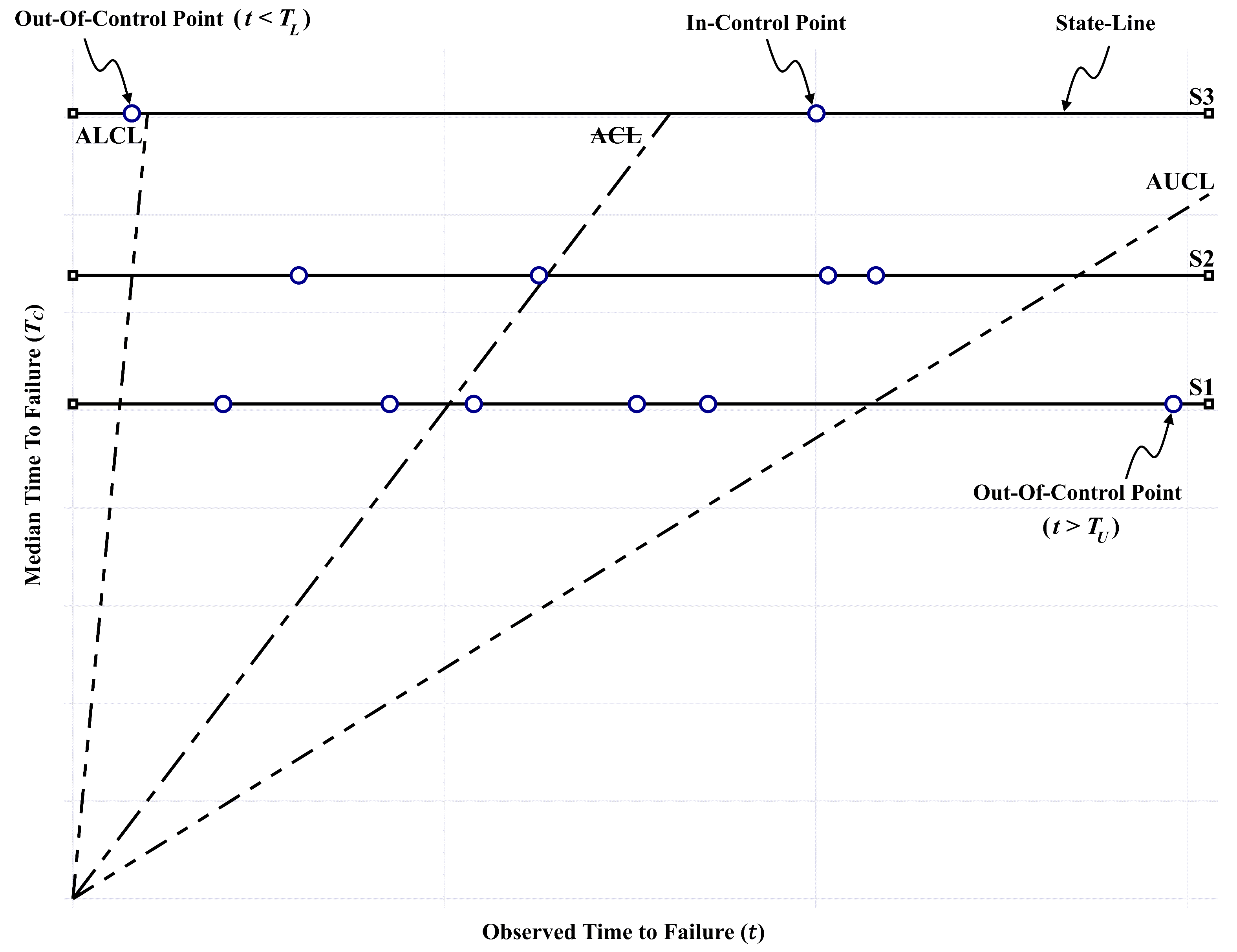}
	\caption{Standard design of the ACC}
	\label{Fig:StandardACC}
\end{figure*}

The ACC is designed so that it depicts each of the system state transitions separately; thus, showing the failure behavior of the whole system as well as the failure behavior of each state. In the ACC (\cref{Fig:StandardACC}), the vertical axis measures the median TTF ($T_C$) of the state transitions. Each state transition is then represented by a horizontal line, called a \emph{state line}, crossing the vertical axis at its corresponding $T_C$. The observed TTF is measured on the horizontal axis.

Each observation is represented by a point that lies on its corresponding state line, and with a horizontal distance, $t$, that represents the observed TTF. Thus, the angle of inclination of the line connecting this point to the origin is determined as
\begin{align}
	\theta = \atan{\frac{T_C}{t}} \label{Eq:theta}
\end{align}
where $\atan{x}$ is the arctangent function (inverse tangent). For simplicity, this angle will be denoted to the observation point. It should be noted that due to using different measures of scale for the horizontal and vertical axes, the appearance of angles may be distorted on the chart.

\subsection{Angular Control Limits}
For convenience and ease of calculations, a function $\rho(a, b)$ is defined here as the \emph{ratio of quantiles} of two values $a$ and $b$,
\begin{align}
	\rho(a, b) = \frac{F^{-1}(a)}{F^{-1}(b)} \label{Eq:rhoFunction}
\end{align}

ACC uses ACLs, each limit with a different angle of inclination on the horizontal axis. The angles of the angular center line (\sout{ACL}), angular lower control limit (ALCL), and angular upper control limit (AUCL) will be denoted $\theta_C$, $\theta_L$, and $\theta_U$, respectively. A suitable value for an acceptable false alarm probability, $c$, should be chosen. A default value of $0.27\%$ is generally used in quality and reliability applications; this value is used throughout this paper.

The time corresponding to each probability control limit can be calculated from the quantile function of the state transition probability distribution. \crefrange{Eq:TC}{Eq:TU} show that the lower probability control limit ($T_L$) is set at a cumulative probability of $c/2$ and the upper probability control limit ($T_U$) is set at a cumulative probability of $1-c/2$. Additionally, for probability limits control charts, the median, rather than the mean, is used as the center line $\paran{T_C=F^{-1}(1/2)}$. This is because, for unsymmetrical probability distributions, the median is the statistic that will ensure a scatter of an equal number of observation points above and below the center line. Consequently, from \cref{Eq:theta}, the angles of the \sout{ACL}, ALCL, and AUCL become
\begin{align}
	\theta_C &= \atan{\frac{T_C}{T_C}} = \atan{\rho(1/2, 1/2)} = 45\degree \label{Eq:thetaC}
\end{align}
\begin{align}
	\theta_L &= \atan{\frac{T_C}{T_L}} = \atan{\rho(1/2, c/2)} \label{Eq:thetaL}
\end{align}
\begin{align}
	\theta_U &= \atan{\frac{T_C}{T_U}} = \atan{\rho(1/2, 1 - c/2)} \label{Eq:thetaU}
\end{align}

It is evident from \cref{Eq:thetaC} that the center line will always have an angle of $45\degree$ regardless of the state transition probability distribution. As presented in \cref{Fig:StandardACC}, an observation point that has an angle of inclination $\theta$ less than $\theta_U$ (point below AUCL), is considered an out-of-control point that represents an increase in the observed TTF. This indicates potential system improvement. Similarly, an observation point that has an angle of inclination $\theta$ greater than $\theta_L$ (point above ALCL), is considered an out-of-control point that represents a decrease in the observed TTF, which is an indication of possible degradation in the system. An in-control observation point will have an angle of inclination $\theta$ such that $\theta_U\leq\theta\leq\theta_L$.

Since $\theta_C$ represents the median of the data, then, if a system is in control, approximately half of the observation points are expected to lie above the center line $\paran{\theta>\theta_C}$, and the other half are expected to lie below it $\paran{\theta<\theta_C}$. This is true for the overall observation points of the system, as well as the observation points on each state line.

\subsection{Exponential ACLs}
For an exponential distribution with a scale parameter (reciprocal of failure rate) $\alpha$, the quantile function is
\begin{align}
	F^{-1}(p) = -\alpha\:\ln{1-p},\quad 0\leq p \leq 1 \label{Eq:quantileExp}
\end{align}

Based on \cref{Eq:quantileExp}, the exponential ratio of quantiles function is defined as
\begin{align}
	\rho(a, b) = \frac{\ln{1 - a}}{\ln{1 - b}} \label{Eq:rhoExp}
\end{align}

By substituting \cref{Eq:rhoExp} in \cref{Eq:thetaL,Eq:thetaU}, the angles of ACLs for the exponential ACC are calculated as
\begin{align}
	\theta_L &= \atan{\frac{\ln{1/2}}{\ln{1 - c/2}}} \approx 89.89\degree \label{Eq:thetaLExp}\\
	\theta_U &= \atan{\frac{\ln{1/2}}{\ln{c/2}}} \approx 5.99\degree \label{Eq:thetaUExp}
\end{align}

\cref{Eq:thetaLExp,Eq:thetaUExp} prove that if all the state transitions are exponentially distributed, then, ACLs do not depend on the scale parameter of these distributions. They depend only on the chosen value of acceptable false alarm probability (taken here to be equal to $0.27\%$, as mentioned before). This enables ACC to monitor all exponentially distributed state transitions using the same ACLs.

\subsection{ACLs with Different Scales}
\cref{Eq:thetaLExp,Eq:thetaUExp} represent the ACLs for an exponential ACC drawn on a linear scale. The value of $\theta_L$ ($89.89\degree$ in this case) becomes very close to the vertical axis. This could limit the chart’s ability in detecting the observations in the region where $\theta>\theta_L$(representing possible system degradation). As a solution, ACCs can be drawn with non-linear scales.

However, the drawing scale must be chosen so that the ACLs remain independent of the distributions’ scale parameters. This is attained if the mapping function of the chosen drawing scale is distributive over division. In other words, the mapping function, $g(x)$, of any chosen scale must satisfy the following condition
\begin{align}
	\frac{g\paran{F^{-1}(a)}}{g\paran{F^{-1}(b)}} = g\paran{\frac{F^{-1}(a)}{F^{-1}(b)}} = g\paran{\rho(a, b)} \label{Eq:distributive}
\end{align}

\cref{Eq:distributive} applies, for instance, to the square and cubic root scales. It does not, however, apply to the logarithmic or exponential scales. See \cref{Tab:thetaForScales} for a list of exponential ACLs using different scales. Using a drawing scale solves the problem of having the lower and upper control limits so close to the vertical and horizontal axes. It should be noted however that it has the disadvantage of having the state lines closer together, which makes the chart more crowded. Thus, choosing a suitable drawing scale is a trade-off between these two factors.

\begin{table}[!h]
	\caption{Angles of Exponential ACLs at different scales}  \label{Tab:thetaForScales}
	\begin{tabularx}{0.48\textwidth}{XXX}
		\toprule
		Scale Type & $\theta_L$ & $\theta_U$ \\
		\midrule

		Linear 		& $89.89\degree$ & $05.99\degree$\\
		Square Root	& $87.47\degree$ & $17.95\degree$\\
		Cubic Root	& $82.88\degree$ & $25.25\degree$\\
		Quartic Root& $78.13\degree$ & $29.64\degree$\\
		\bottomrule
	\end{tabularx}
\end{table}

\section{Generalized Design of ACC} \label{Sec:GeneralACC}
The main purpose of generalization is to allow using the ACC design to monitor MSSs having state transitions with any non-negative continuous probability distribution in addition to the exponential distribution. \cref{Eq:thetaLExp,Eq:thetaUExp} determine the ACLs when all state transitions are exponentially distributed. These control limits are independent of the exponential distribution's scale parameter. (Notice that the exponential distribution doesn't have a shape parameter.)

However, for state transitions with probability distributions that do have both scale and shape parameters, ACLs depend only on the shape parameter as will be shown later in \crefrange{Sec:WeibullACLs}{Sec:GammaACLs} and as summarized in \cref{Tab:Summary} in \cref{Sec:Summary}. It can also be shown that ACLs would be  dependent on the location parameter as well if the three-parameter versions of these distributions were applied. Therefore, the design of the ACC has to be modified to allow for such dependency.

\begin{figure*}[!hbt]
	\centering
	\includegraphics[width=0.8\linewidth]{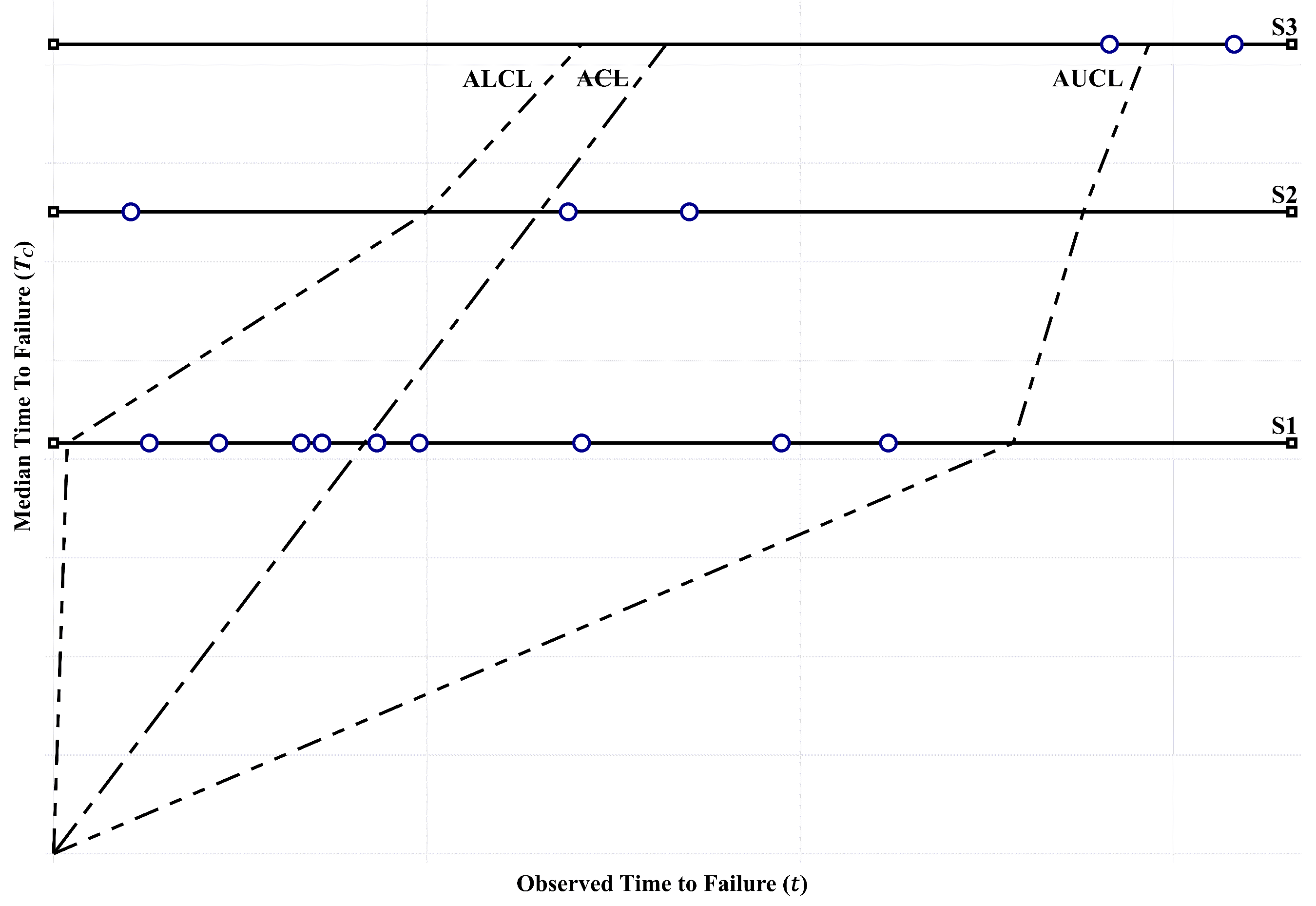}
	\caption{Generalized design of ACC}
	\label{Fig:GeneralACC}
\end{figure*}

The generalized ACC (\cref{Fig:GeneralACC}) allows the ACLs to zigzag throw state lines crossing each state line at its corresponding $T_C$, $T_L$, and $T_U$. In other words, each ACL composes of several line segments, one for each state. The angle of inclination of each segment will be calculated based on its corresponding state transition probability distribution.

Next in \crefrange{Sec:WeibullACLs}{Sec:GammaACLs}, the angles of ACLs are calculated for state transitions of Weibull, lognormal, Fr\'echet, and gamma distributions, respectively. The calculations are based on the linear scale but can be easily modified to other scales. For more information about these distributions, their parameters, and their quantile functions, refer to \citet{Bury.1999}, \citet{Forbes.2011}, and \citet{OConnor.2011}.

\subsection{Weibull ACLs} \label{Sec:WeibullACLs}
The quantile function of Weibull distribution with a scale parameter $\alpha$ and a shape parameter $\beta$ is estimated as
\begin{align}
	F^{-1}(p) = \alpha \paran{-\ln{1-p}}^{\frac{1}{\beta}},\quad 0\leq p \leq 1 \label{Eq:quantileWeibull}
\end{align}

From \cref{Eq:quantileWeibull}, the Weibull ratio of quantiles function becomes
\begin{align}
	\rho(a, b) = \frac{\alpha \paran{-\ln{1-a}}^{\frac{1}{\beta}}}{\alpha \paran{-\ln{1-b}}^{\frac{1}{\beta}}} = \paran{\frac{\ln{1-a}}{\ln{1-b}}}^{\frac{1}{\beta}} \label{Eq:rhoWeibull}
\end{align}

Using \cref{Eq:rhoWeibull} with \cref{Eq:thetaL,Eq:thetaU}, the Weibull ACLs' angles become
\begin{align}
	\theta_L &= \atan{\paran{\frac{\ln{1/2}}{\ln{1 - c/2}}}^{\frac{1}{\beta}}} \nonumber \\
	&\approx \atan{\sqrt[\beta]{513.096}} \label{Eq:thetaLWeibull}
\end{align}
\begin{align}
	\theta_U &= \atan{\paran{\frac{\ln{1/2}}{\ln{c/2}}}^{\frac{1}{\beta}}} \nonumber \\
	&\approx \atan{\sqrt[\beta]{0.105}} \label{Eq:thetaUWeibull}
\end{align}

\begin{figure*}[!hbt]
	\centering

	\begin{subfigure}{0.45\textwidth}
		\centering
		\includegraphics[width=\linewidth]{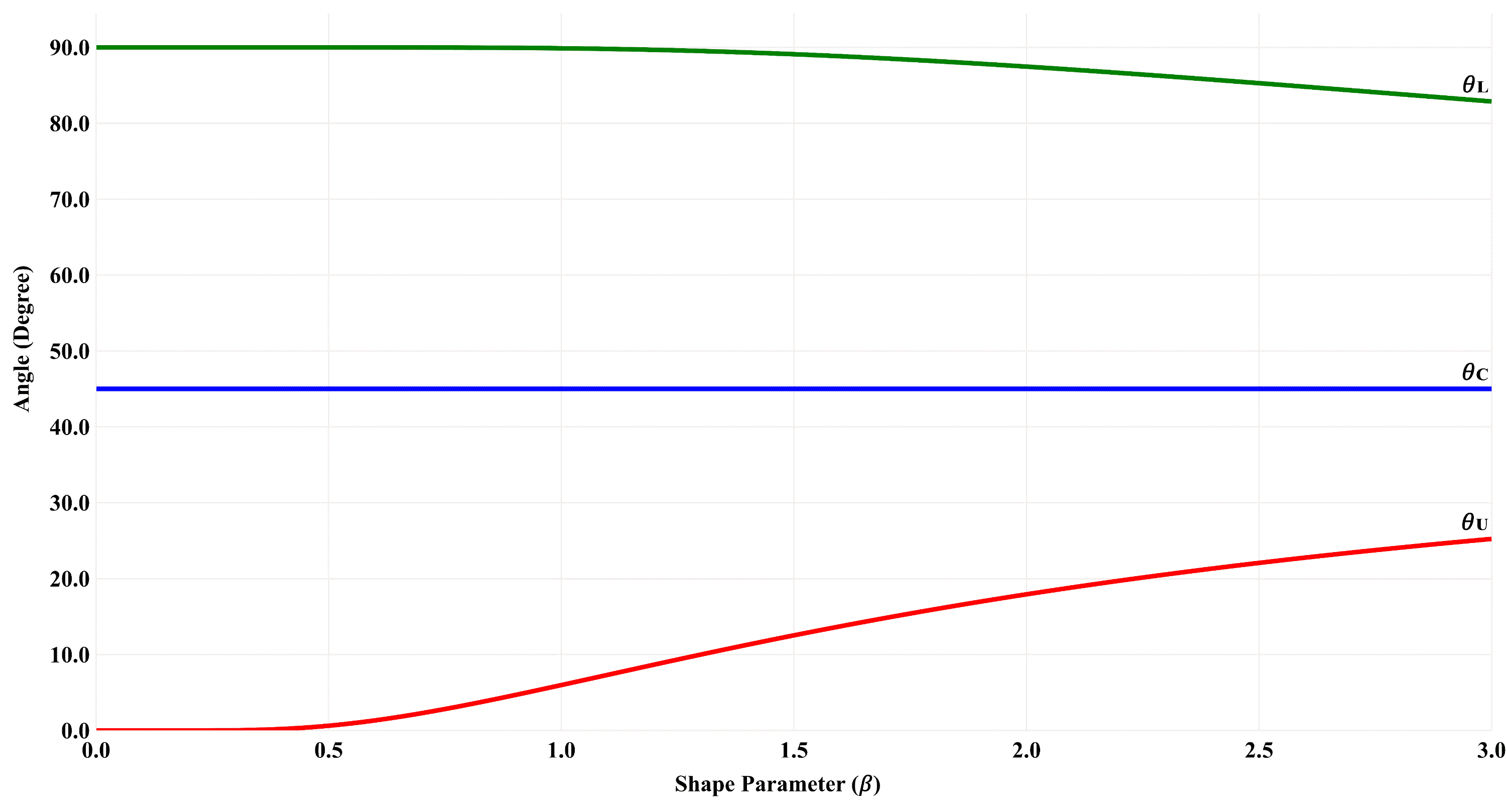}
		\caption{Linear scale}
	\end{subfigure}
	\hspace{1em}
	\begin{subfigure}{0.45\textwidth}
		\centering
		\includegraphics[width=\linewidth]{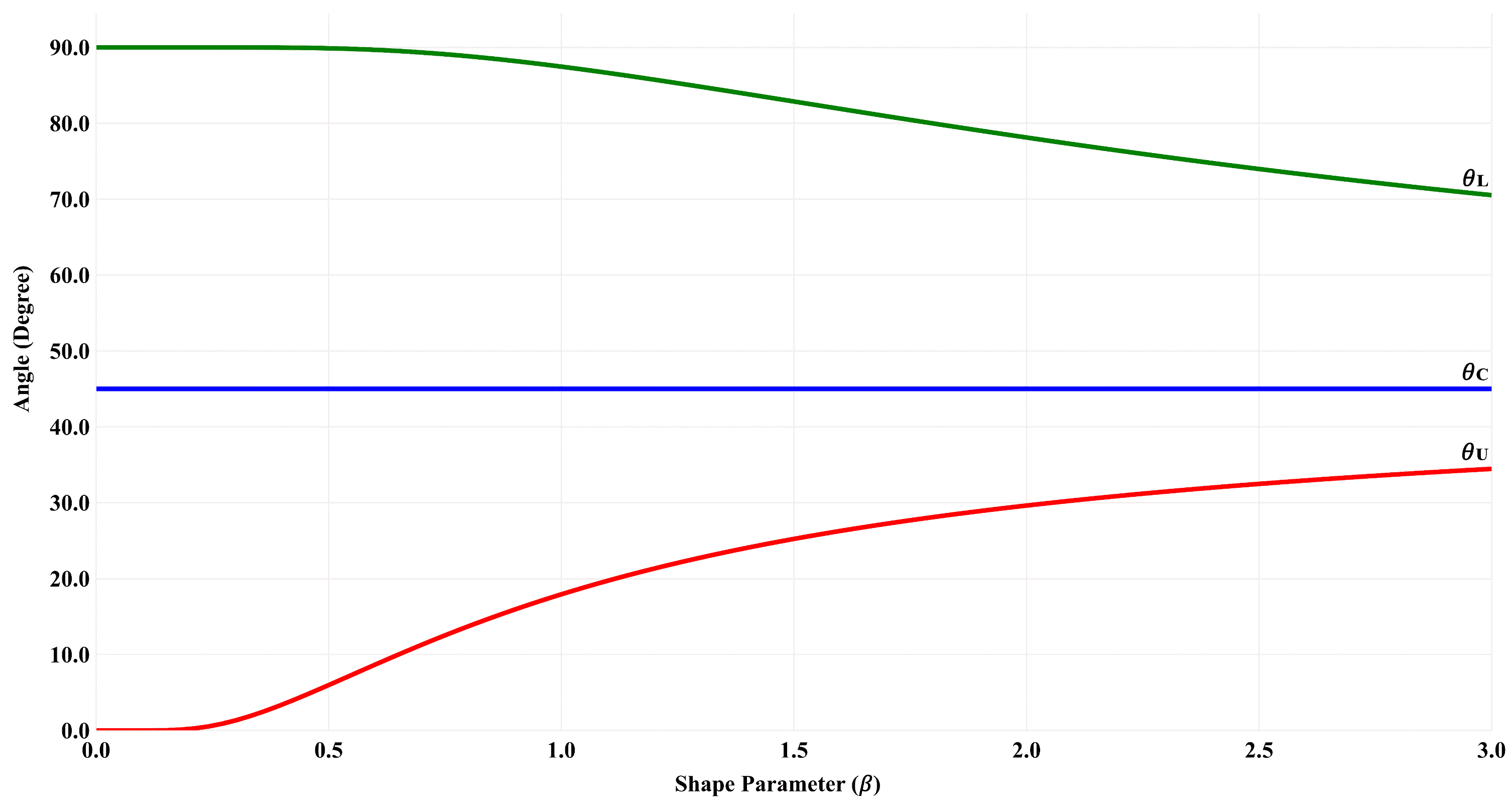}
		\caption{Square root scale}
	\end{subfigure}

	\begin{subfigure}{0.45\textwidth}
		\centering
		\includegraphics[width=\linewidth]{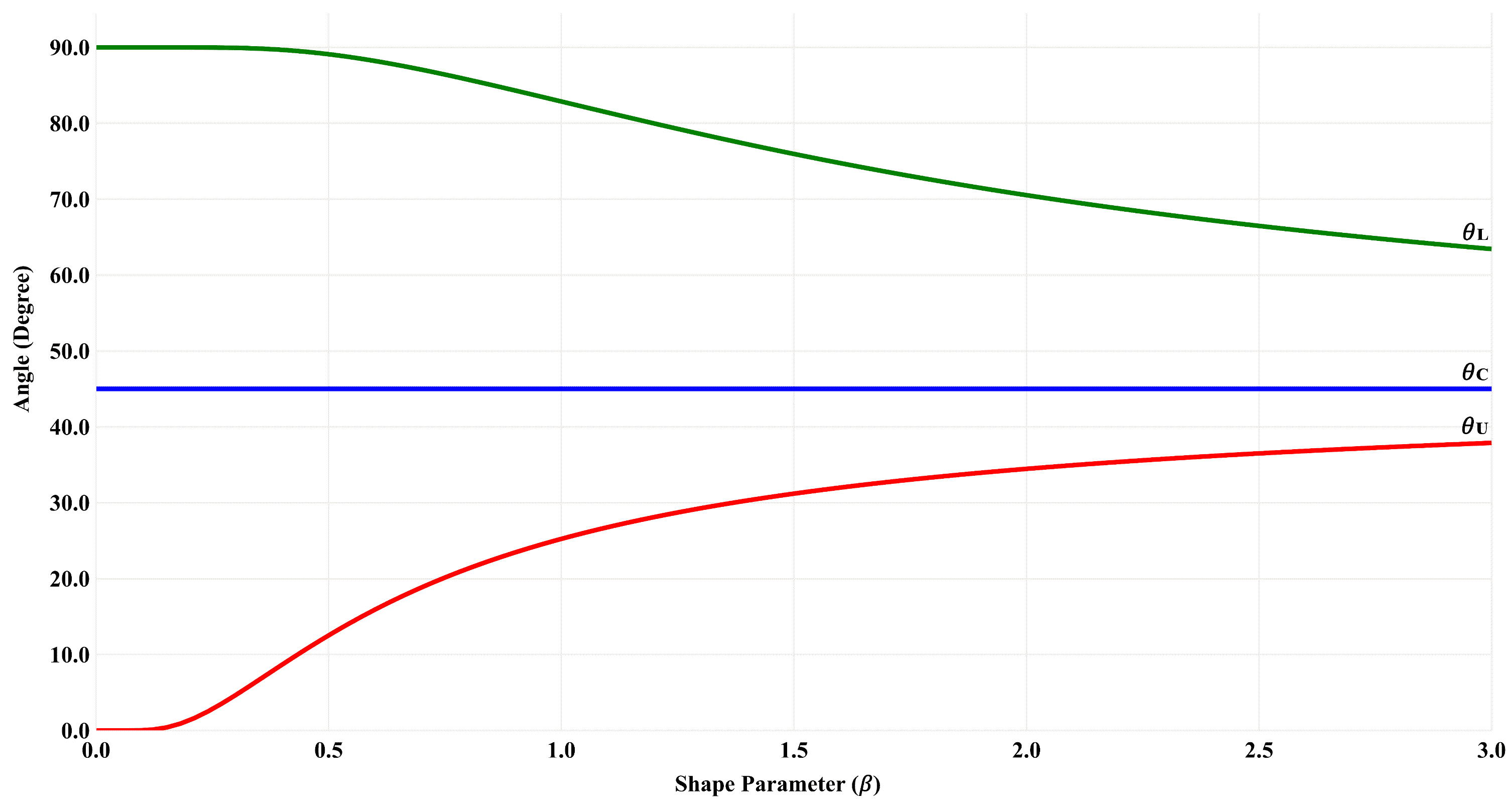}
		\caption{Cubic root scale}
	\end{subfigure}
	\hspace{1em}
	\begin{subfigure}{0.45\textwidth}
		\centering
		\includegraphics[width=\linewidth]{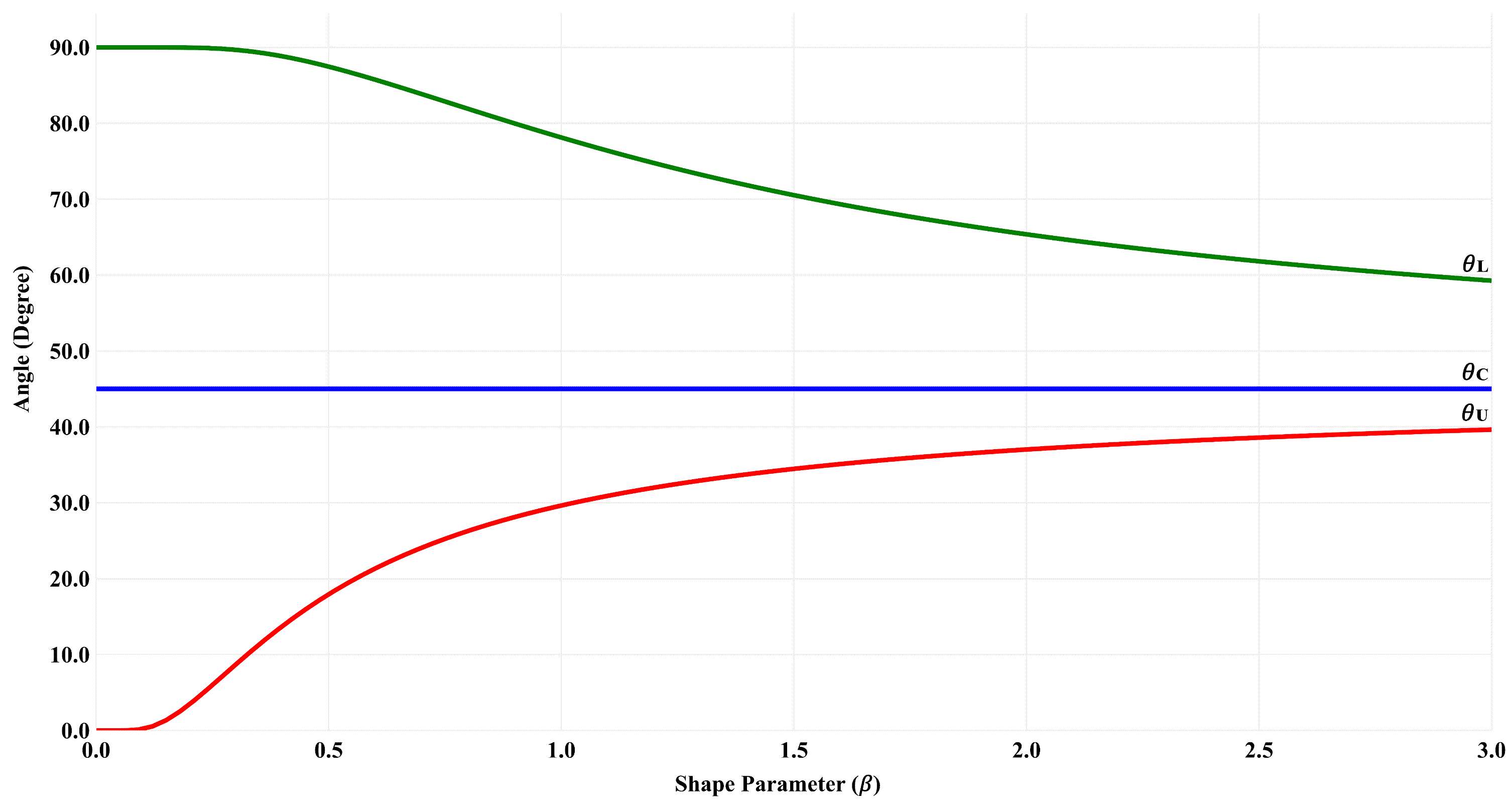}

		\caption{Quartic root scale}
	\end{subfigure}

	\caption{Relationship between Weibull's ACLs and the shape parameter $\beta$}
	\label{Fig:weibullACLs}
\end{figure*}

\cref{Eq:thetaLWeibull,Eq:thetaUWeibull} show that the ACLs, in the case of the Weibull distribution, are dependent upon the shape parameter. \cref{Fig:weibullACLs} demonstrates the relationships between the Weibull ACLs and the shape parameter $\beta$ for different drawing scales.

\cref{Eq:thetaLExp,Eq:thetaUExp} are directly related to \cref{Eq:thetaLWeibull,Eq:thetaUWeibull} as the exponential distribution is a special case of Weibull distribution when the shape parameter $\beta = 1$. Similarly, Rayleigh ACLs can be easily calculated based on the fact that Rayleigh distribution is a special case of Weibull distribution when the shape parameter $\beta = 2$ (See \cref{Tab:Summary} in \cref{Sec:Summary}).

\subsection{Lognormal ACLs} \label{Sec:LognormalACLs}
For the lognormal distribution, quantile and ratio of quantiles functions are estimated as

\begin{align}
	F^{-1}(p) &= \exp\paran{\alpha + \beta\,\Phi^{-1}(p)},\quad 0\leq p \leq 1 \label{Eq:quantileLognormal}\\
	\rho(a, b) &= \exp\paran{\beta\paran{\Phi^{-1}(a) - \Phi^{-1}(b)}} \label{Eq:rhoLognormal}
\end{align}
where $\alpha$ is the scale parameter (normal distribution mean), $\beta$ (normal distribution standard deviation) is the shape parameter, and $\Phi^{-1}(p)$ is the standard normal quantile function. Applying \cref{Eq:rhoLognormal} in \cref{Eq:thetaL,Eq:thetaU} the angles of ACLs for the lognormal distribution become
\begin{align}
	\vspace*{-2\baselineskip}
	\theta_L &= \atan{\exp\paran{-\beta\,\Phi^{-1}(c/2)}} \nonumber\\
	&= \atan{\exp\paran{3\beta}} \label{Eq:thetaLLognormal}
\end{align}
\begin{align}
	\theta_U &= \atan{\exp\paran{-\beta\,\Phi^{-1}(1 - c/2)}} \nonumber\\
	&= \atan{\exp\paran{-3\beta}} \label{Eq:thetaULognormal}
\end{align}

\begin{figure*}[!hbt]
	\centering

	\begin{subfigure}{0.45\textwidth}
		\centering
		\includegraphics[width=\linewidth]{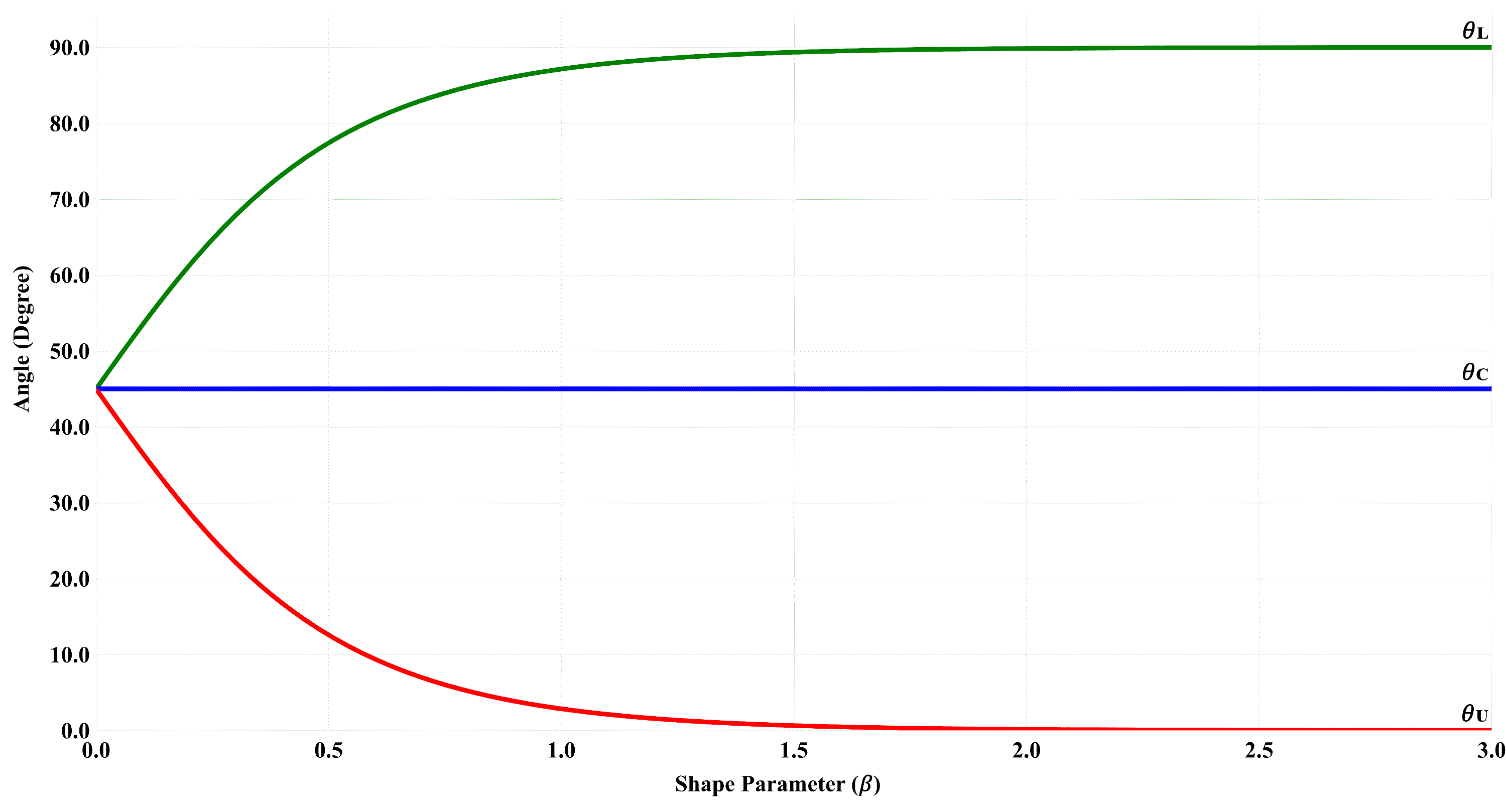}
		\caption{Linear scale}
	\end{subfigure}
	\hspace{1em}
	\begin{subfigure}{0.45\textwidth}
		\centering
		\includegraphics[width=\linewidth]{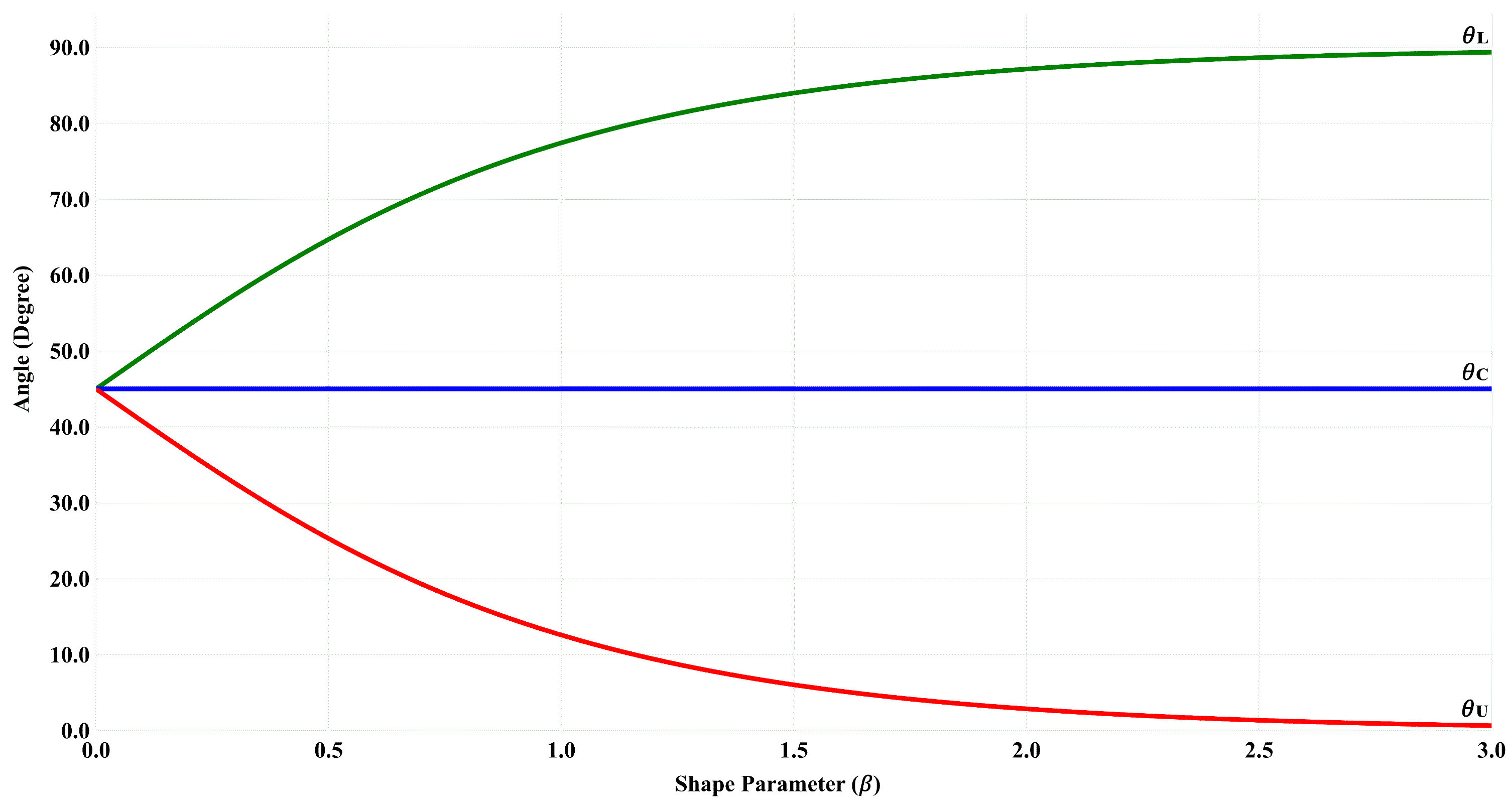}
		\caption{Square root scale}
	\end{subfigure}

	\begin{subfigure}{0.45\textwidth}
		\centering
		\includegraphics[width=\linewidth]{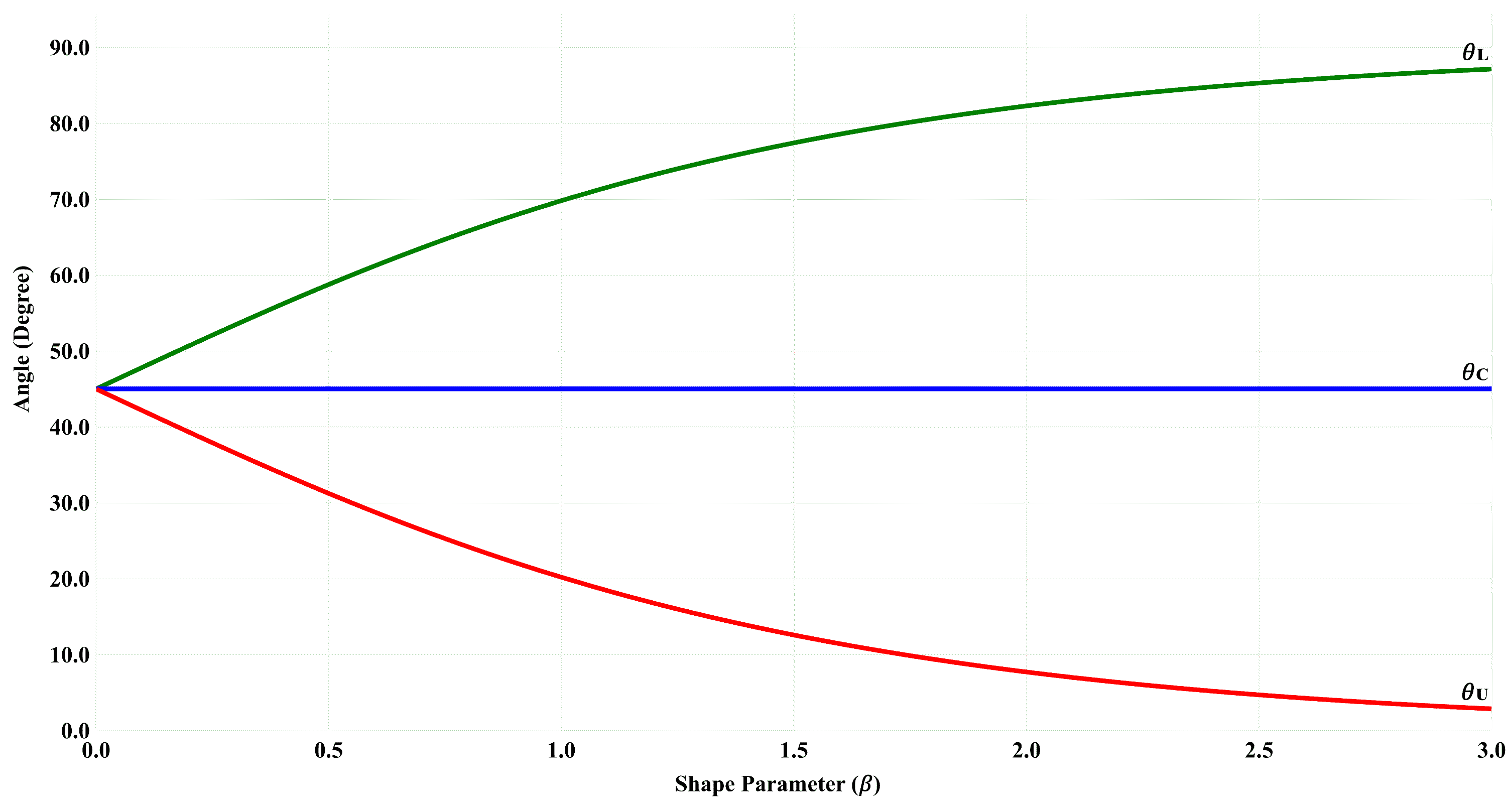}
		\caption{Cubic root scale}
	\end{subfigure}
	\hspace{1em}
	\begin{subfigure}{0.45\textwidth}
		\centering
		\includegraphics[width=\linewidth]{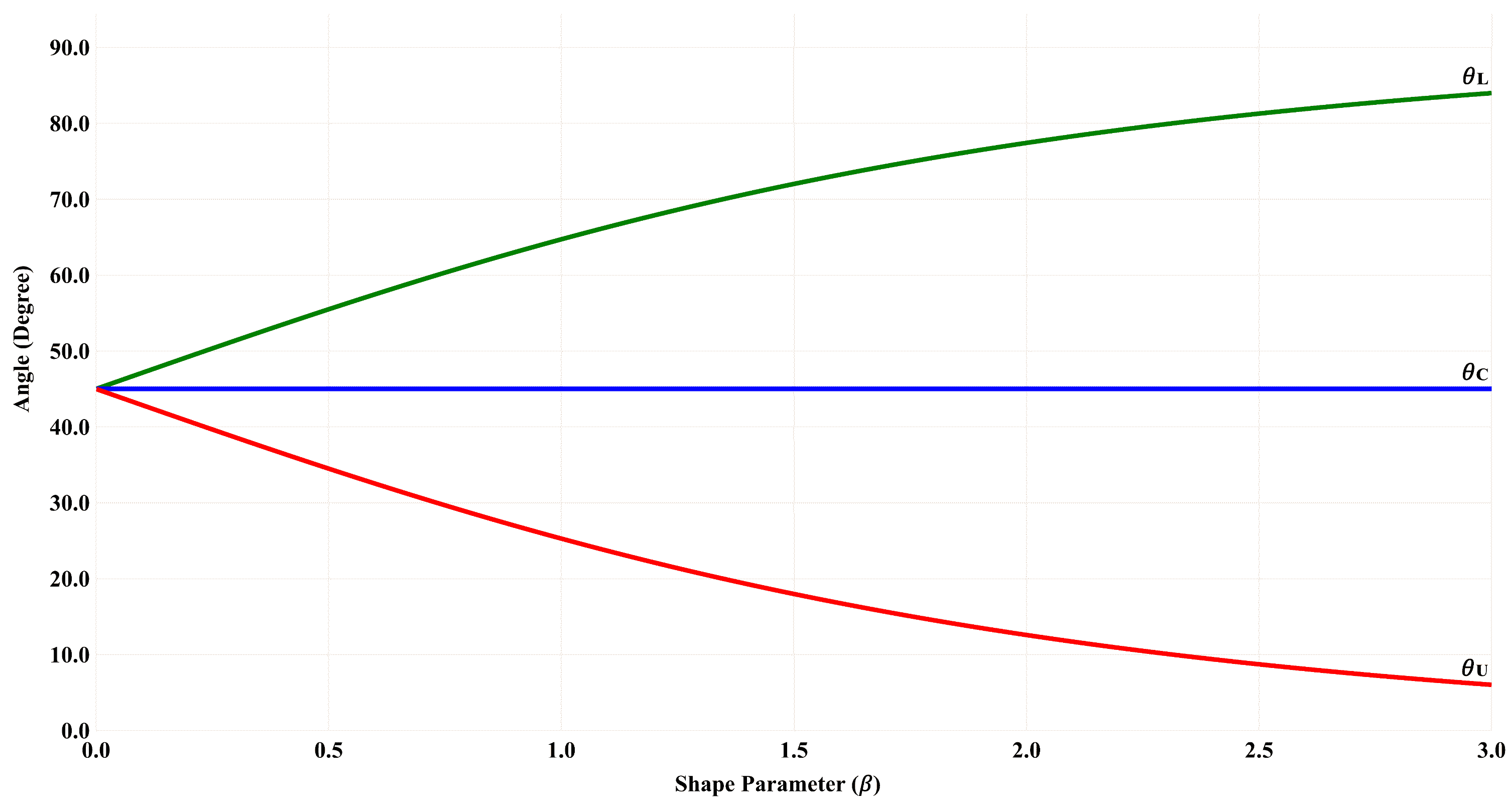}
		\caption{Quartic root scale}
	\end{subfigure}

	\caption{Relationship between lognormal ACLs and the shape parameter $\beta$}
	\label{Fig:lognormalACLs}
\end{figure*}

Note that $\Phi^{-1}(1/2) = 0$. Moreover, when the value of $c$ is taken as the standard value of $0.27\%$ corresponding to a $3$-sigma control limit, then $\Phi^{-1}(c/2)=-3$ and $\Phi^{-1}(1 - c/2)=3$. The relationship between the lognormal ACLs and the shape parameter $\beta$ is shown in \cref{Fig:lognormalACLs} for different drawing scales.

\subsection{Fr\'echet ACLs} \label{Sec:FrechetACLs}
The quantile and ratio of quantiles functions for the Fr\'echet distribution (inverse Weibull distribution) are defined as
\begin{align}
	F^{-1}(p) &= \alpha\paran{-\ln p}^{-\frac{1}{\beta}},\quad 0\leq p \leq 1 \label{Eq:quantileFrechet} \\
	\rho(a, b) &= \paran{\frac{\ln b}{\ln a}}^{\frac{1}{\beta}} \label{Eq:rhoFrechet}
\end{align}

\begin{figure*}[!hbt]
	\centering

	\begin{subfigure}{0.45\textwidth}
		\centering
		\includegraphics[width=\linewidth]{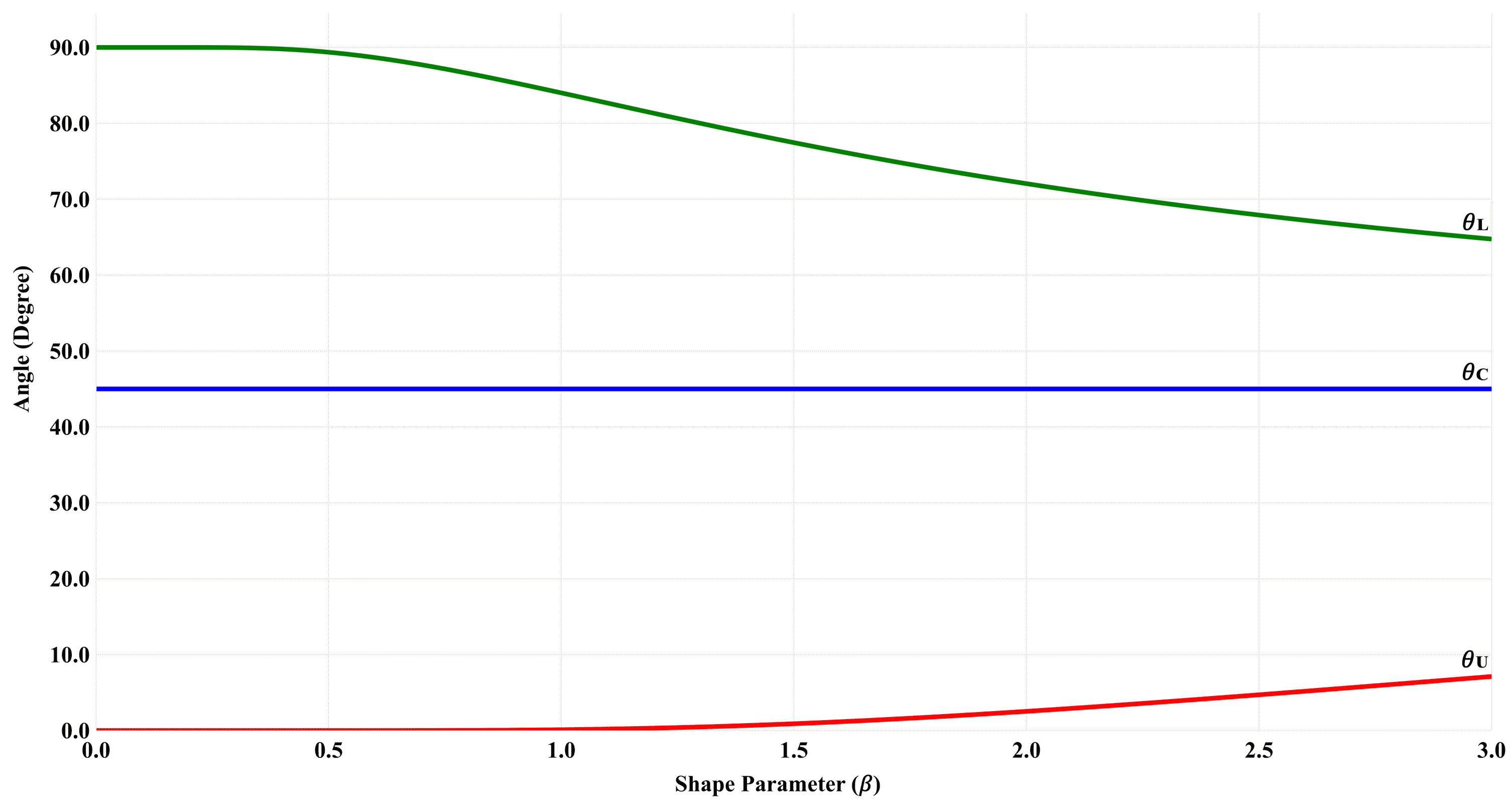}
		\caption{Linear scale}
	\end{subfigure}
	\hspace{1em}
	\begin{subfigure}{0.45\textwidth}
		\centering
		\includegraphics[width=\linewidth]{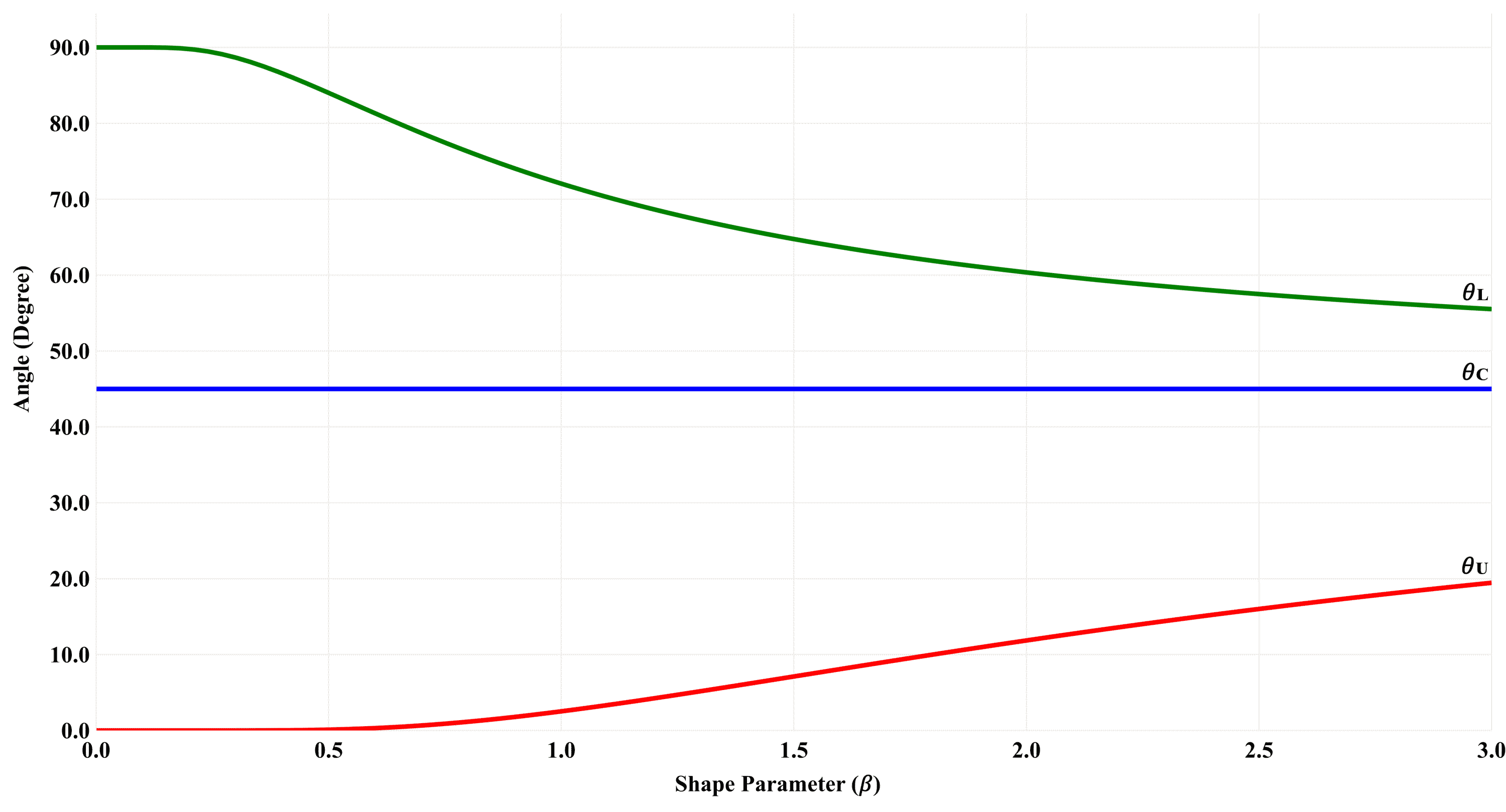}
		\caption{Square root scale}
	\end{subfigure}

	\begin{subfigure}{0.45\textwidth}
		\centering
		\includegraphics[width=\linewidth]{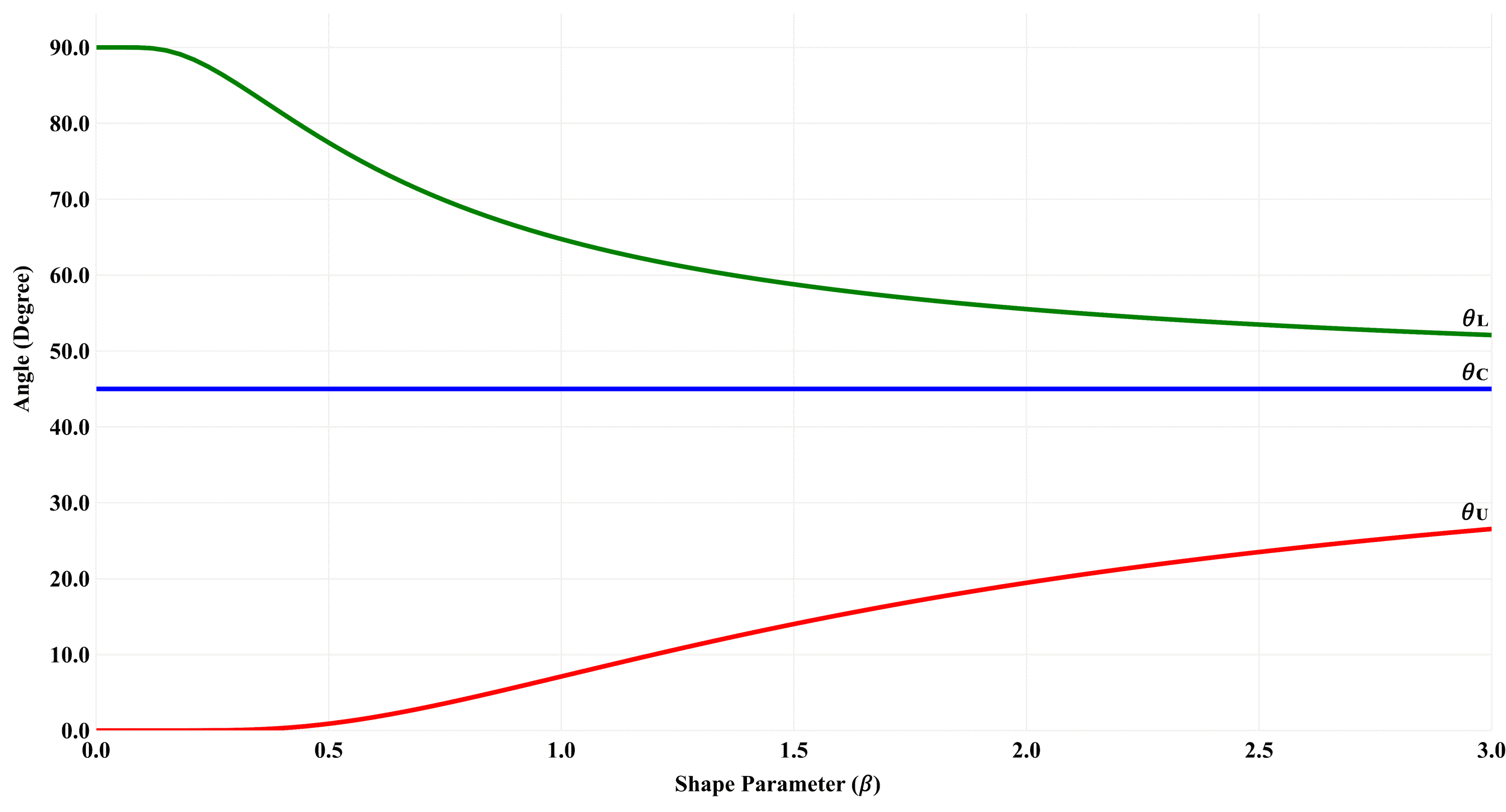}
		\caption{Cubic root scale}
	\end{subfigure}
	\hspace{1em}
	\begin{subfigure}{0.45\textwidth}
		\centering
		\includegraphics[width=\linewidth]{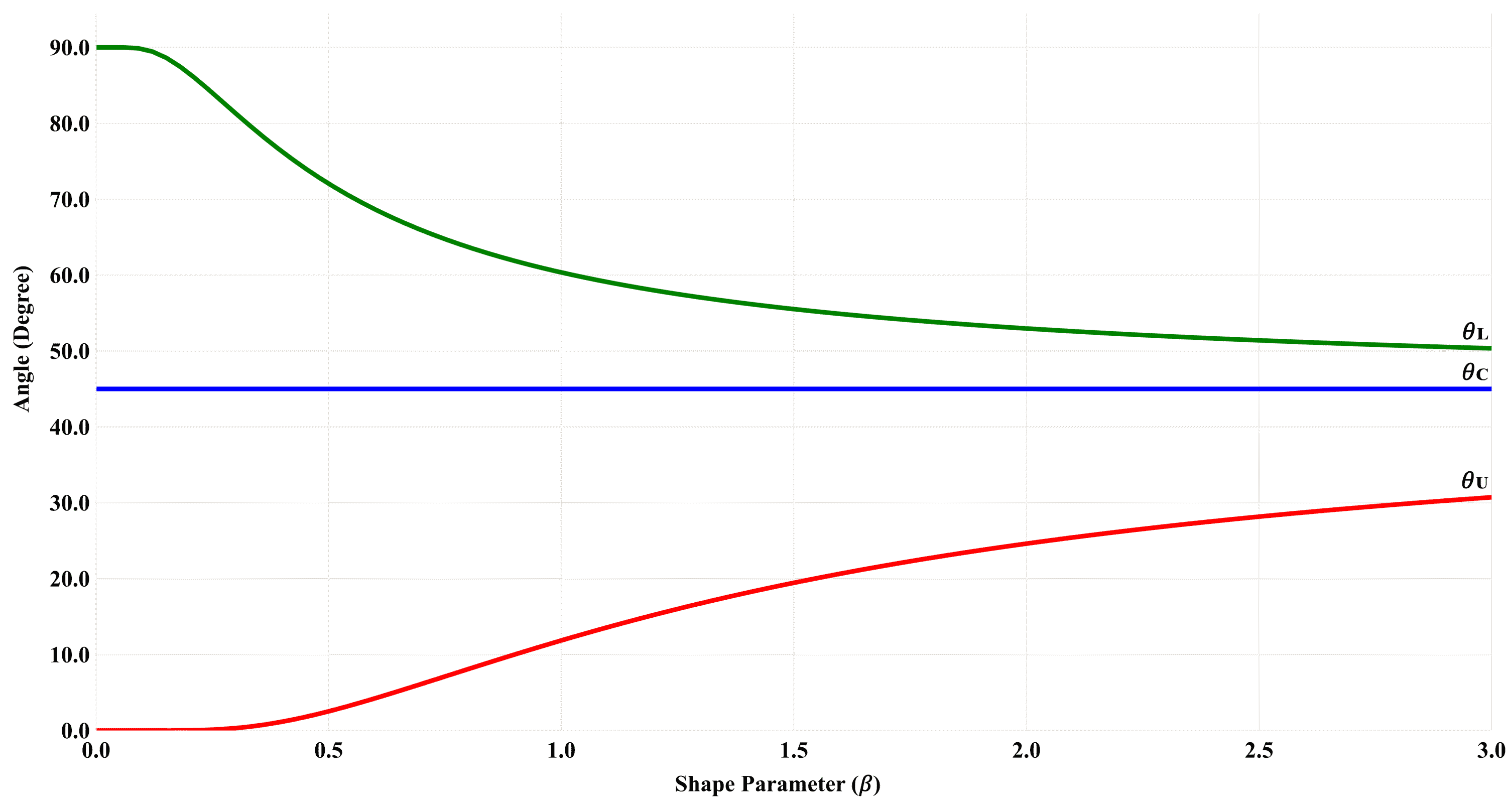}
		\caption{Quartic root scale}
	\end{subfigure}

	\caption{Relationship between Fr\'echet's ACLs and the shape parameter $\beta$}
	\label{Fig:frechetACLs}
\end{figure*}

Using \cref{Eq:rhoFrechet} with \cref{Eq:thetaL,Eq:thetaU} gives the ACLs for the Fr\'echet distribution as demonstrated in \cref{Eq:thetaLFrechet,Eq:thetaUFrechet}. \cref{Fig:frechetACLs} demonstrates the relationship between the angles of Fr\'echet's ACLs and the shape parameter $\beta$.
\begin{align}
	\theta_L &= \atan{\paran{\frac{\ln{c/2}}{\ln{1/2}}}^{\frac{1}{\beta}}} \nonumber \\
	&\approx \atan{\sqrt[\beta]{9.533}} \label{Eq:thetaLFrechet}\\
	\theta_U &= \atan{\paran{\frac{\ln{1-c/2}}{\ln{1/2}}}^{\frac{1}{\beta}}} \nonumber\\
	&\approx \atan{\sqrt[\beta]{0.002}} \label{Eq:thetaUFrechet}
\end{align}

\subsection{Gamma ACLs} \label{Sec:GammaACLs}
\begin{figure*}[!hbt]
	\centering

	\begin{subfigure}{0.45\textwidth}
		\centering
		\includegraphics[width=\linewidth]{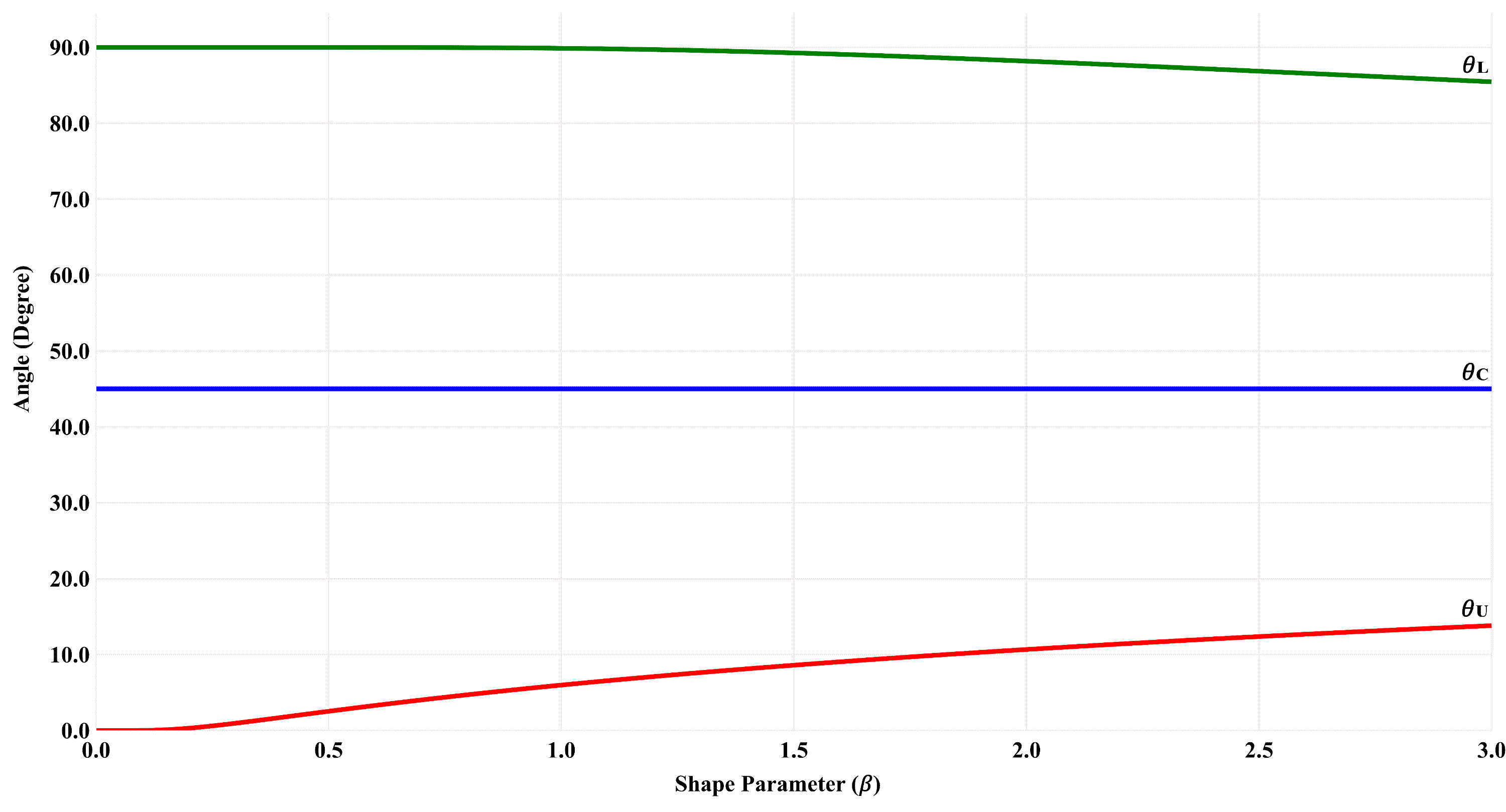}
		\caption{Linear scale}
	\end{subfigure}
	\hspace{1em}
	\begin{subfigure}{0.45\textwidth}
		\centering
		\includegraphics[width=\linewidth]{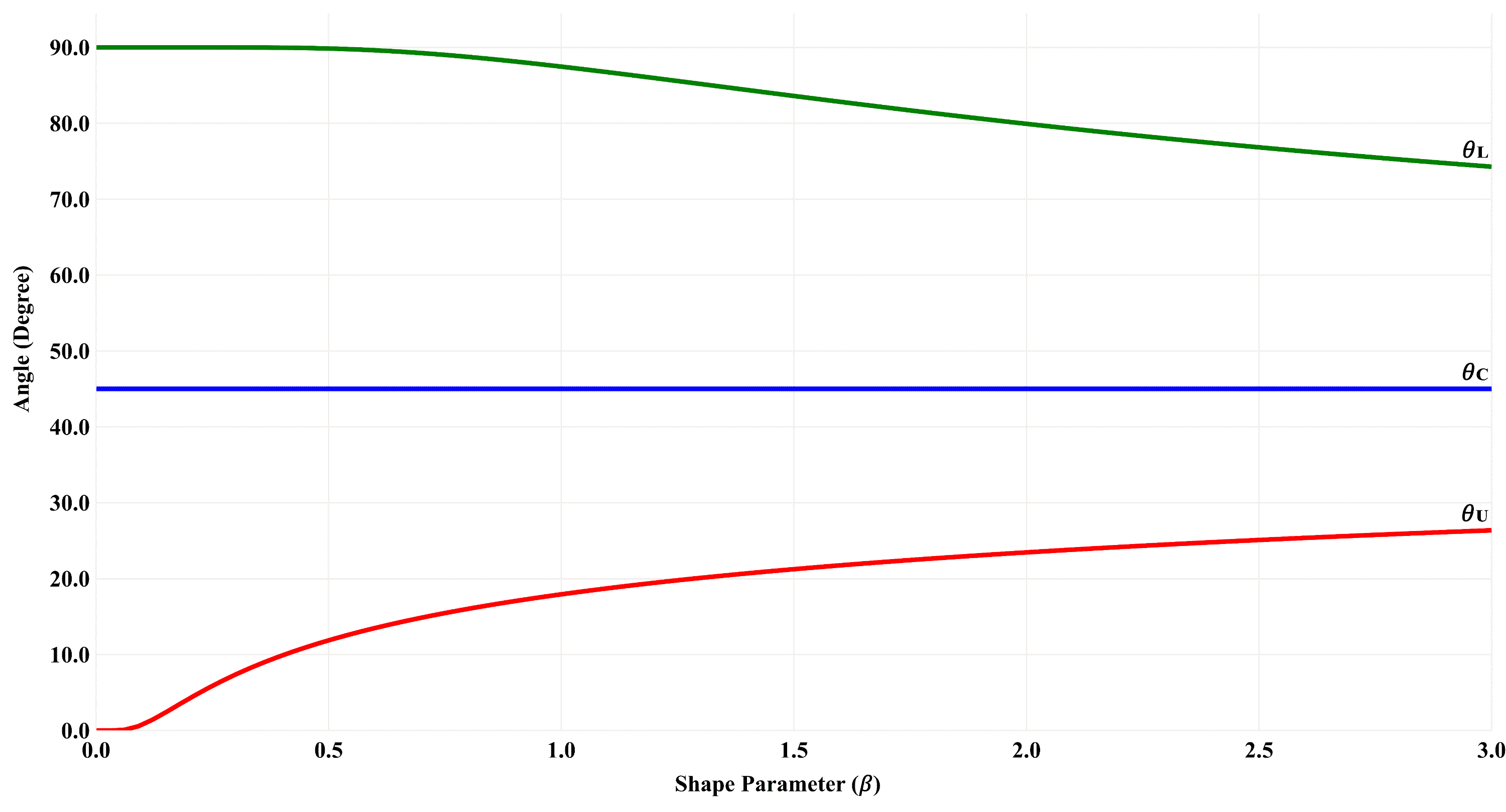}
		\caption{Square root scale}
	\end{subfigure}

	\begin{subfigure}{0.45\textwidth}
		\centering
		\includegraphics[width=\linewidth]{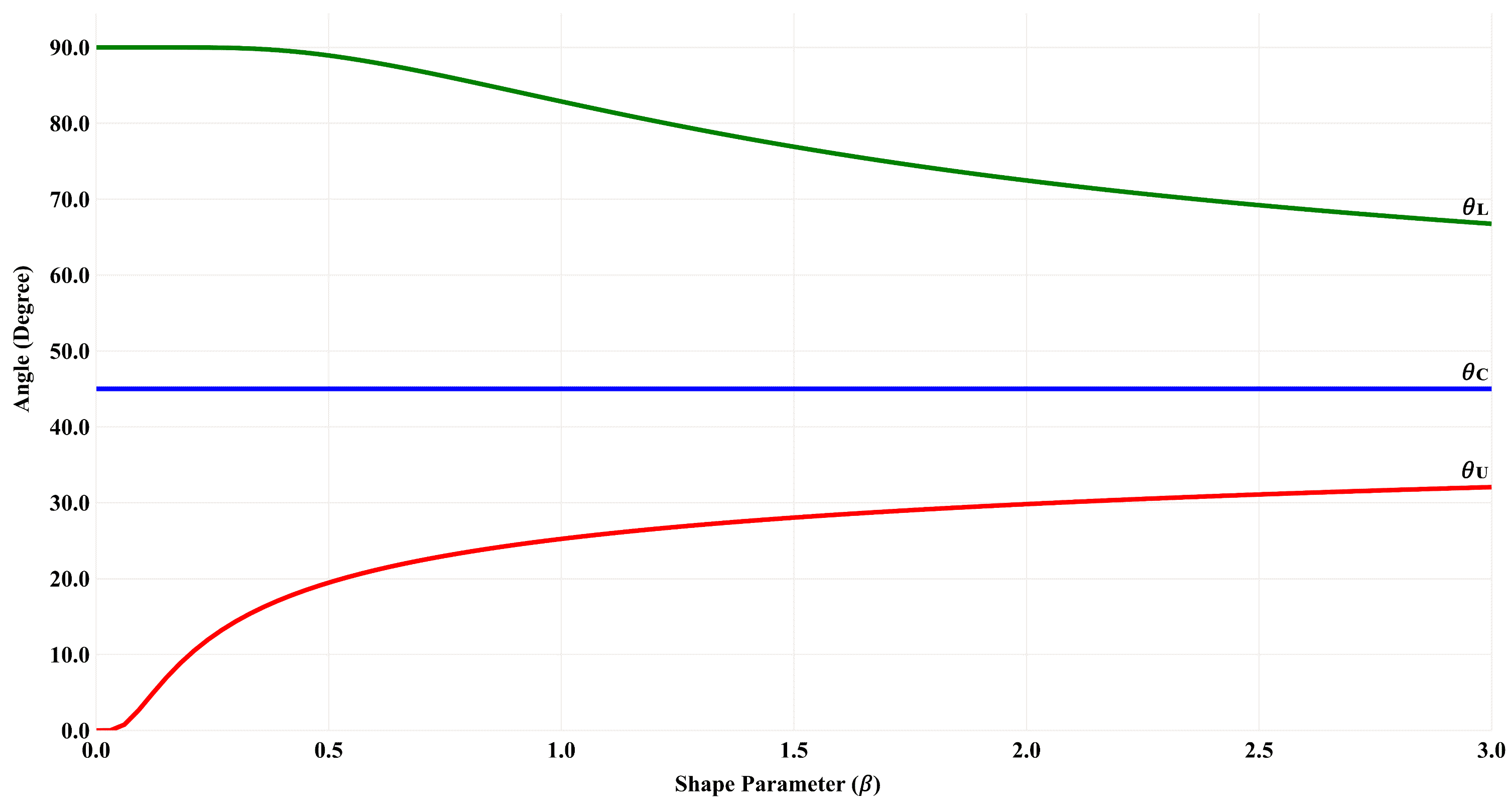}
		\caption{Cubic root scale}
	\end{subfigure}
	\hspace{1em}
	\begin{subfigure}{0.45\textwidth}
		\centering
		\includegraphics[width=\linewidth]{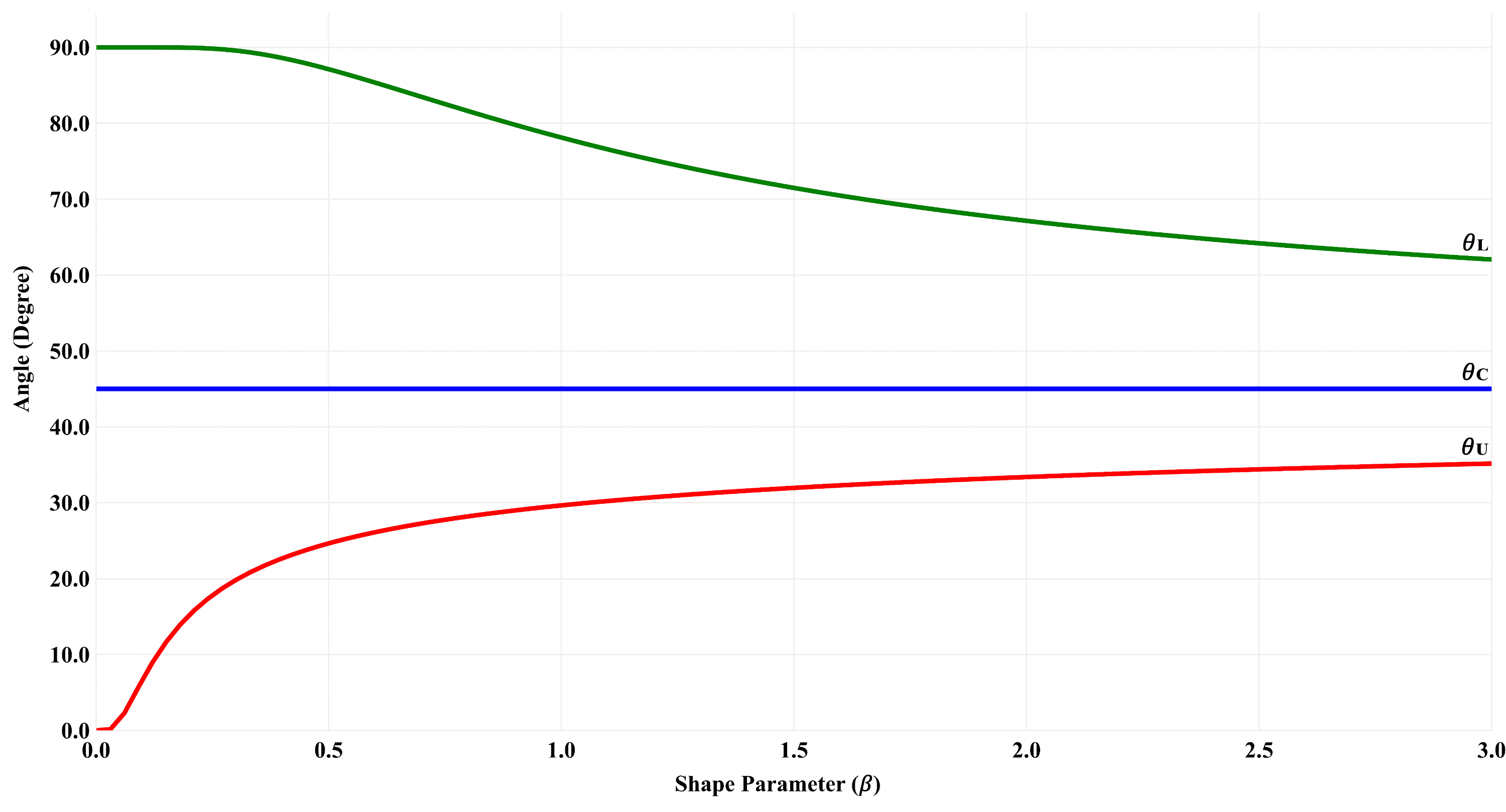}
		\caption{Quartic root scale}
	\end{subfigure}

	\caption{Relationship between gamma's ACLs and the shape parameter $\beta$}
	\label{Fig:gammaACLs}
\end{figure*}

The cumulative probability function of the gamma distribution is given by
\begin{align}
	F(t) &= \frac{1}{\Gamma(\beta)}\int_{0}^{t/\alpha} z^{\beta-1}\exp(-z)\:dz,\quad t > 0 \label{Eq:gammaCDF}
\end{align}
where $\alpha$ is the scale parameter, $\beta$ is the shape parameter, and $\Gamma(x)$ is the gamma function. ACLs for gamma distribution are calculated numerically as the gamma distribution does not have a simple closed-form for the quantile function. The numerical calculations proved that the angles of ACLs in the case of the gamma distribution depend on the shape parameter only and not on the scale parameter. \cref{Fig:gammaACLs} shows the relationship between ACLs and shape parameter $\beta$ of the gamma distribution.

The same calculations can be followed for Erlang distribution as a special case of gamma distribution in which the shape parameter can have only positive integer values.

\section{Illustrative Examples} \label{Sec:Examples}
Three simulated examples are presented to illustrate the implementation of the standard and generalized ACCs. The results are generated using the \emph{AC Charts Software} \citep{Janada.2022}. The cubic root drawing scale is adopted in all charts.

\subsection*{Example I}
\begin{figure*}[!h]
	\centering
	\includegraphics[width=0.73\linewidth]{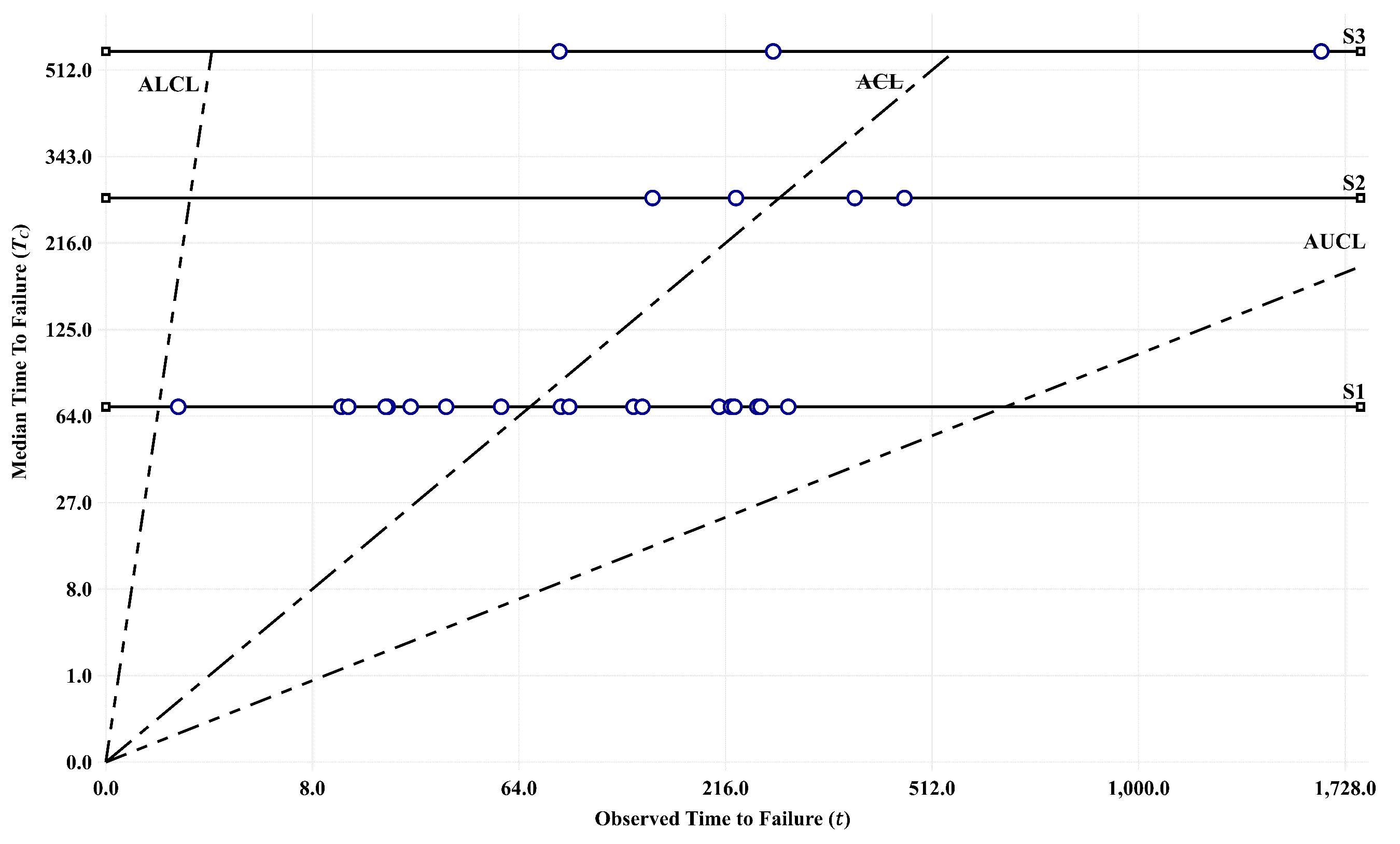}
	\caption{ACC for the first 25 observations of Example I}
	\label{Fig:Ex1Before}
\end{figure*}

\begin{figure*}[!h]
	\centering
	\includegraphics[width=0.73\linewidth]{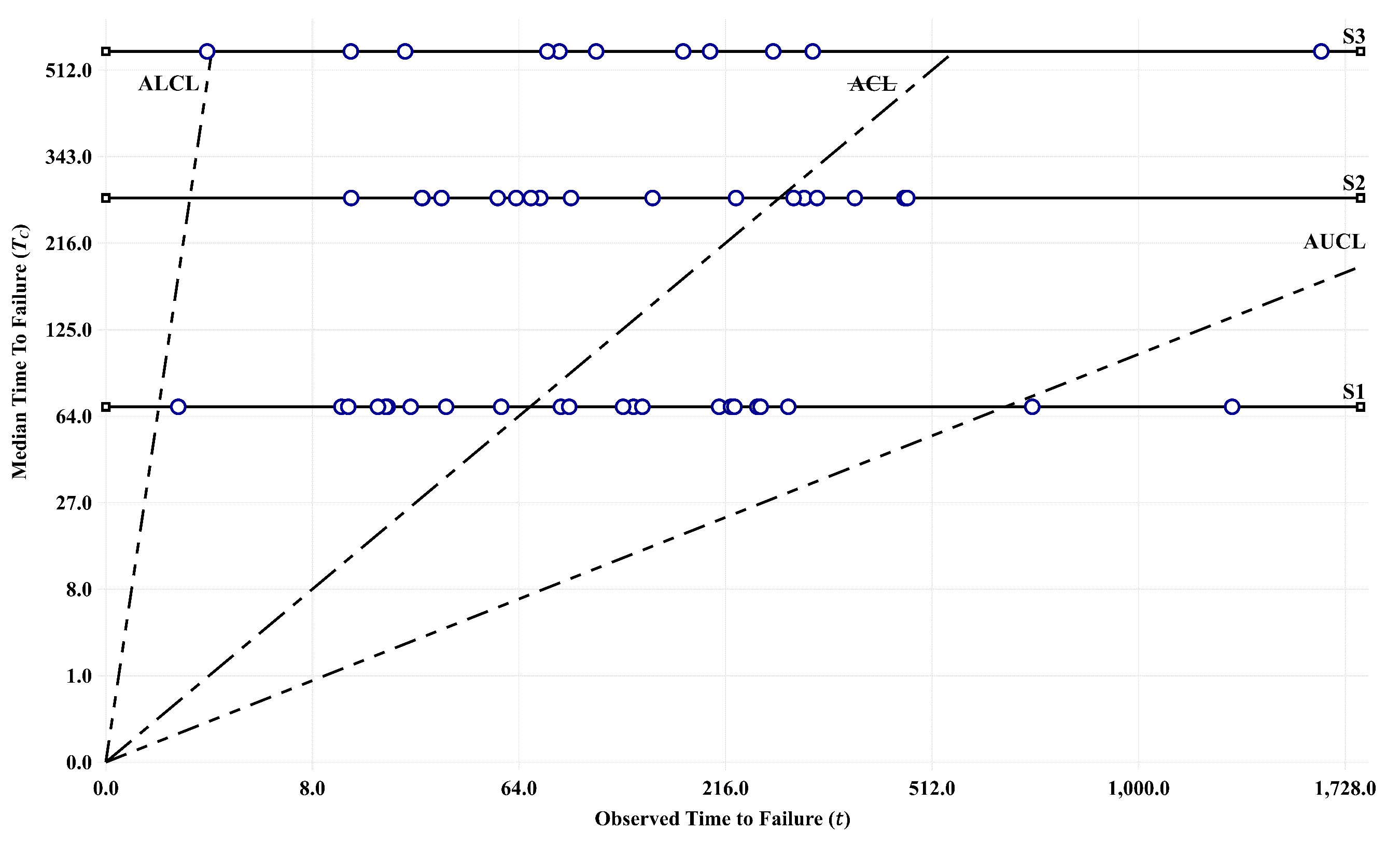}
	\caption{ACC for all 50 observations of Example I}
	\label{Fig:Ex1After}
\end{figure*}

Consider an MSS that has four possible performance levels—a perfect state and three reduced performance states. All state transitions are modeled by an exponential distribution, with a scale parameter of 100, 400, and 800, for S$_1$, S$_2$, and S$_3$, respectively. \cref{Tab:Example1Data} lists the simulated TTFs for Example I. The first 25 times are randomly generated from the exponential distribution using the same scale parameters of the corresponding state transitions. To simulate some shifts in the failure behavior of the system, another 25 times are randomly generated from the exponential distribution with scale parameters of 400, 200, and 200, for state transitions S$_1$, S$_2$, and S$_3$, respectively. This simulation supposes that an improvement occurs with S$_1$, and degradation occurs with S$_2$ and S$_3$. \cref{Fig:Ex1Before} shows the ACC for only the first 25 TTFs and \cref{Fig:Ex1After} shows the ACC for the entire 50 TTFs.

\begin{table}[!h]
	\caption{Simulated TTFs for Example I}
	\label{Tab:Example1Data}
	\begin{tabularx}{0.47\textwidth}{llX|llX}
		\toprule
		No. & State & TTF & No. & State & TTF\\
		\midrule
		1  & 1 & 288.50  & 26 & 1 & 125.56  \\
		2  & 1 & 133.33  & 27 & 1 & 721.89  \\
		3  & 1 & 56.09   & 28 & 2 & 34.34   \\
		4  & 1 & 251.05  & 29 & 2 & 309.10  \\
		5  & 1 & 35.75   & 30 & 1 & 18.28   \\
		6  & 2 & 462.16  & 31 & 2 & 467.33  \\
		7  & 1 & 209.36  & 32 & 3 & 320.57  \\
		8  & 1 & 85.49   & 33 & 1 & 1296.80 \\
		9  & 2 & 381.33  & 34 & 2 & 74.42   \\
		10 & 1 & 11.85   & 35 & 2 & 91.36   \\
		11 & 1 & 140.19  & 36 & 3 & 174.59  \\
		12 & 2 & 227.15  & 37 & 2 & 13.39   \\
		13 & 1 & 254.75  & 38 & 2 & 326.59  \\
		14 & 1 & 20.31   & 39 & 2 & 28.80   \\
		15 & 3 & 269.73  & 40 & 3 & 107.00  \\
		16 & 2 & 148.37  & 41 & 2 & 295.25  \\
		17 & 1 & 12.93   & 42 & 3 & 0.94    \\
		18 & 1 & 25.72   & 43 & 2 & 28.69   \\
		19 & 1 & 221.74  & 44 & 3 & 13.35   \\
		20 & 1 & 19.88   & 45 & 2 & 62.67   \\
		21 & 3 & 84.72   & 46 & 3 & 78.12   \\
		22 & 3 & 1629.98 & 47 & 2 & 69.69   \\
		23 & 1 & 90.28   & 48 & 3 & 200.30  \\
		24 & 1 & 225.47  & 49 & 3 & 24.33   \\
		25 & 1 & 0.35    & 50 & 2 & 54.60   \\
		\bottomrule
	\end{tabularx}
\end{table}

\cref{Fig:Ex1Before} shows that the observation points are, to a great extent, equally distributed below and above the \sout{ACL}. This is true for the overall system (12 points above \sout{ACL} and 13 below), as well as it is for each state line. This indicates an in-control behavior of the observed MSS. Contrary, \cref{Fig:Ex1After} clearly shows an out-of-control behavior of the system. For S$_1$, there are two points below the AUCL indicating improvement of the system at this particular state transition. Conversely, for S$_3$, the point above the ALCL indicates possible system degradation at this state transition. For S$_2$, there are no out-of-control points. However, an out-of-control behavior can still be detected as nearly two-thirds of the points on S$_2$ (11 out of 17) fall above the \sout{ACL}.

\subsection*{Example II}
\begin{figure*}[!h]
	\centering
	\includegraphics[width=0.73\linewidth]{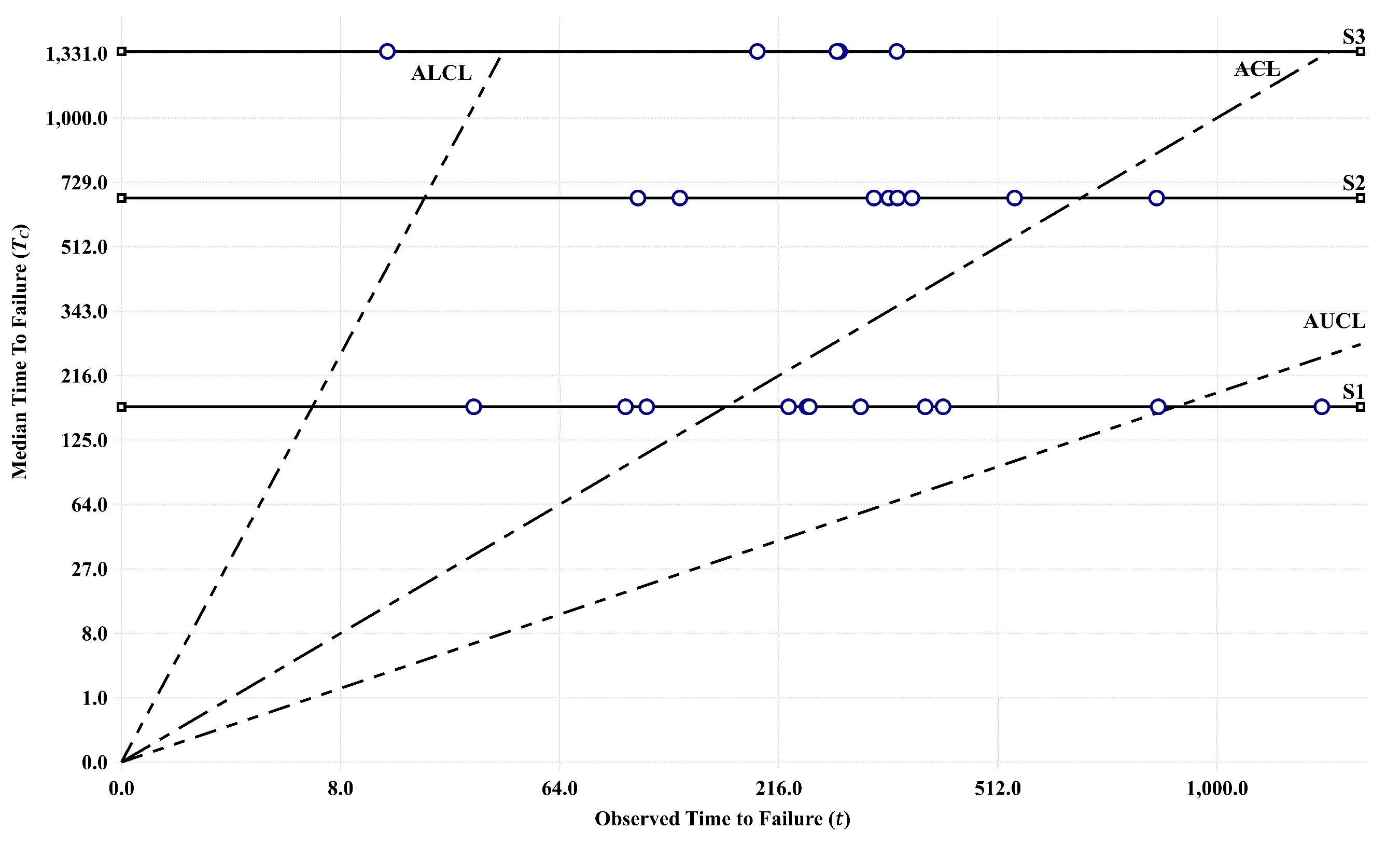}
	\caption{ACC for Example II}
	\label{Fig:Ex2}
\end{figure*}

The MSS described in Example I is considered. However, the ACC will be used to monitor the cumulative TTF between every two failures (of the same state) as in the case of the $t_r$-chart. In \cref{Tab:Example2Data}, the cumulative sum of TTFs every two failures is calculated from the original data in \cref{Tab:Example1Data}. For this application, each state transition will be modeled with Erlang distribution with a shape parameter $\beta = 2$ (as it is the sum of two exponential distributions). The scale parameters of these Erlang distributions are the same as their exponential counterparts in Example I. Since the three state transitions have the same shape parameter and ACLs are dependent only on the shape parameter, then the ACLs' angles will not differ between the three state transitions. The ACC of this example is shown in \cref{Fig:Ex2}.

\begin{table}[!h]
	\caption{Cumulative TTFs for Example II}
	\label{Tab:Example2Data}
	\begin{tabularx}{0.47\textwidth}{llX|llX}
		\toprule
		No. & State & TTF & No. & State & TTF\\
		\midrule
		1  & 1 & 421.83 & 13 & 1 & 847.45  \\
		2  & 1 & 307.14 & 14 & 2 & 343.44  \\
		3  & 1 & 245.11 & 15 & 3 & 1950.55 \\
		4  & 2 & 843.49 & 16 & 1 & 1315.08 \\
		5  & 1 & 97.34  & 17 & 2 & 541.75  \\
		6  & 1 & 394.94 & 18 & 2 & 104.75  \\
		7  & 2 & 375.52 & 19 & 2 & 355.39  \\
		8  & 1 & 33.24  & 20 & 3 & 281.59  \\
		9  & 1 & 247.46 & 21 & 2 & 323.94  \\
		10 & 3 & 354.45 & 22 & 3 & 14.29   \\
		11 & 1 & 110.16 & 23 & 2 & 132.36  \\
		12 & 1 & 225.82 & 24 & 3 & 278.42  \\
		\bottomrule
	\end{tabularx}
\end{table}

The analysis followed in Example I can be also followed in this example. However, the current example produces only one out-of-control point on S$_1$ rather than two out-of-control points as in Example I. This could indicate a decrease in the chart's sensitivity to detect out-of-control points.

\subsection*{Example III}
\begin{figure*}[!h]
	\centering
	\includegraphics[width=0.73\linewidth]{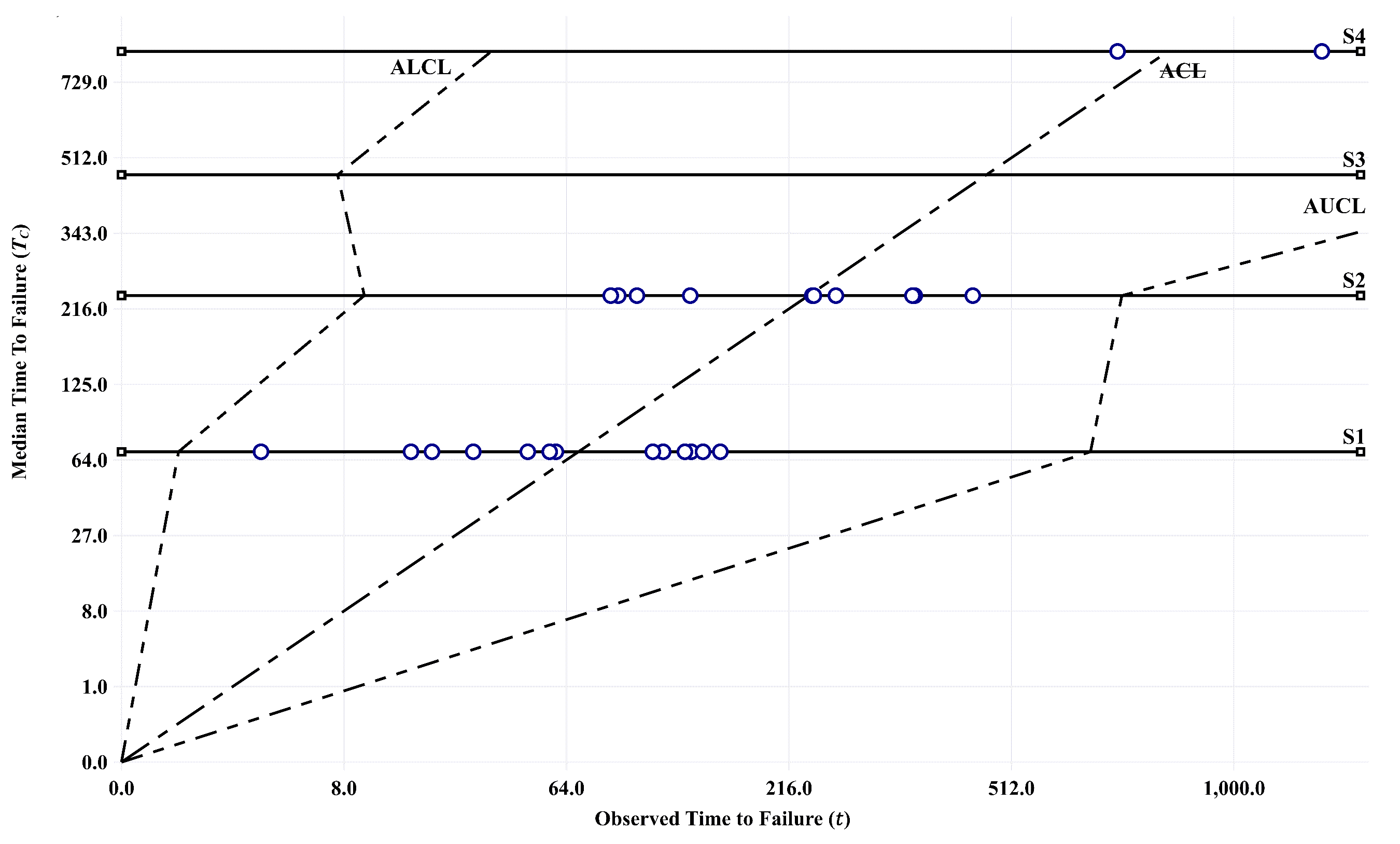}
	\caption{ACC for the first 25 observations of Example III}
	\label{Fig:Ex3Before}
\end{figure*}

\begin{figure*}[!h]
	\centering
	\includegraphics[width=0.73\linewidth]{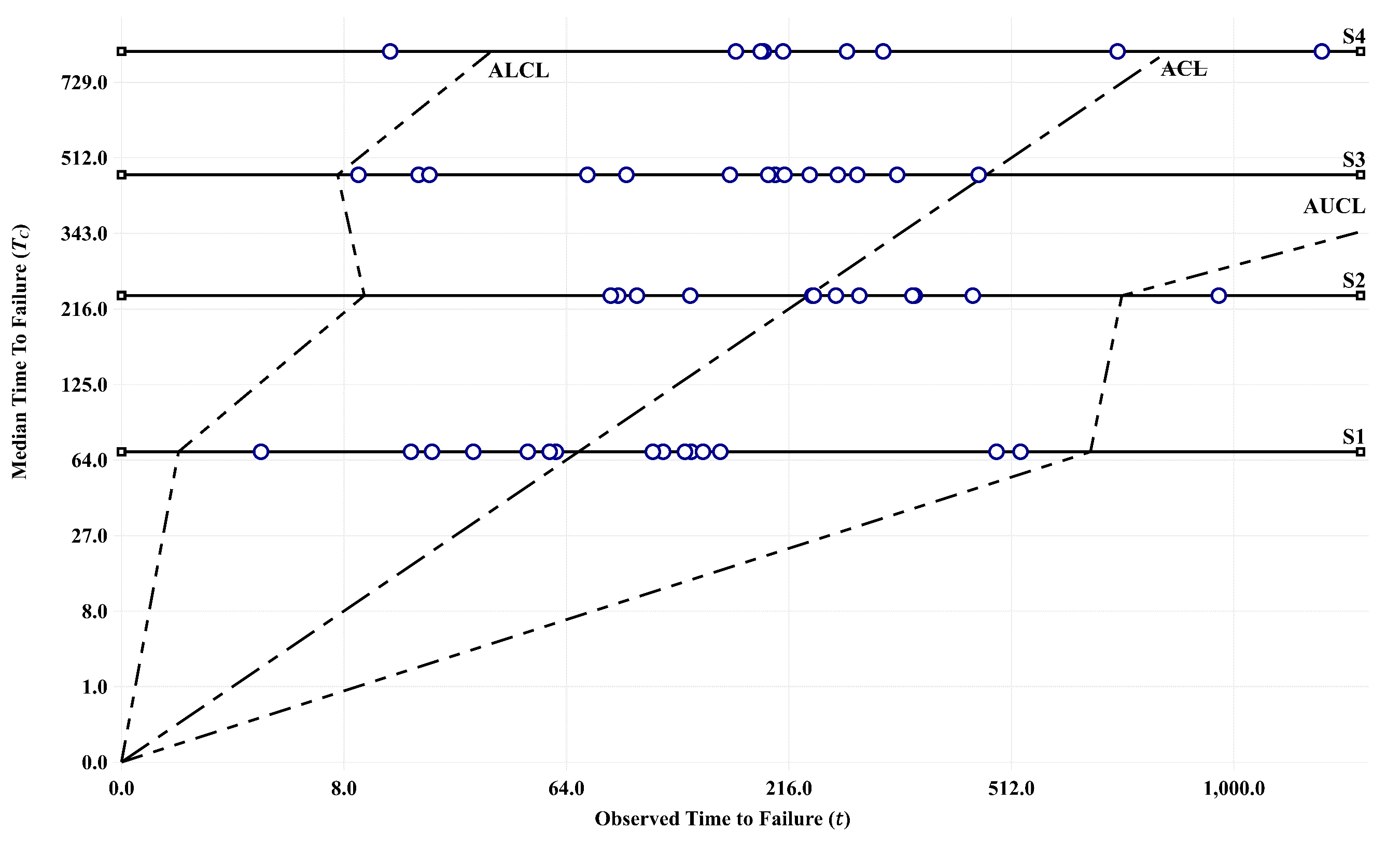}
	\caption{ACC for all 50 observations of Example III}
	\label{Fig:Ex3After}
\end{figure*}

The MSS is considered to have four possible state transitions. The corresponding probability distributions are summarized in \cref{Tab:Example3Info} (note that Rayleigh distribution is a Weibull distribution with $\beta = 2$). As in Example I, the first 25 simulated TTFs are randomly generated using the same probability distributions of the corresponding state transitions and then shown on the ACC in \cref{Fig:Ex3Before}. For the next 25 TTFs, the scale parameters are changed to 300, 300, 200, and 200, for S$_1$, S$_2$, S$_3$, and S$_4$, respectively, while the shape parameters are kept the same. \cref{Fig:Ex3After} shows the ACC for the entire 50 TTFs.

\begin{table}[!h]
	\caption{State Transitions of the MSS in Example III}
	\label{Tab:Example3Info}
	\begin{tabularx}{0.47\textwidth}{XXXX}
		\toprule
		State Transition & Probability Distribution & Scale Parameter & Shape Parameter\\
		\midrule
		S$_1$ & Gamma 		& 100  & 1.0 \\
		S$_2$ & Rayleigh 	& 200  & --- \\
		S$_3$ & Weibull 	& 600  & 1.5 \\
		S$_4$ & Weibull 	& 1000 & 2.0 \\
		\bottomrule
	\end{tabularx}
\end{table}

\cref{Fig:Ex3After} clearly shows the out-of-control behavior of the system's state transitions. In addition to analyzing the failure behavior of each-state transition, the chart can be used to monitor the overall failure behavior of the MSS. In \cref{Fig:Ex3Before}, the observation points were equally distributed about the \sout{ACL} (12 points above \sout{ACL} and 13 below). But referring to \cref{Fig:Ex3After}, it is found that two-thirds of the observation points (33 out of 50) lie above \sout{ACL}. This indicates an overall improvement in the reliability of the system.

\section{Summary} \label{Sec:Summary}
\cref{Tab:Summary} summarizes the values/formulas of ACLs' angles for different state transition probability distributions, using a linear drawing scale (with $c=0.27\%$). The calculations of ACLs’ angles for distributions that do not have a closed-form for the quantile function (e.g., gamma distribution) can be also completed numerically. In addition, \cref{Tab:Comparison} compares the proposed designs of ACCs.

\begin{table*}[!h]
	\caption{Values/formulas of ACLs' angles for different distributions}  \label{Tab:Summary}
	\begin{tabularx}{1\textwidth}{XXX}
		\toprule
		State Distribution & $\theta_L$ & $\theta_L$\\
		\midrule
		Exponential & $89.89\degree$	& $05.99\degree$\\
		Rayleigh 	& $87.47\degree$	& $17.95\degree$\\
		Weibull
		& $\atan{\sqrt[\beta]{513.096}}$
		& $\atan{\sqrt[\beta]{0.105}}$\\
		Fr\'echet
		& $\atan{\sqrt[\beta]{1/0.105}}$
		& $\atan{\sqrt[\beta]{1/513.096}}$\\
		Lognormal
		& $\atan{\exp(3\beta)}$
		& $\atan{\exp(-3\beta)}$\\
		\bottomrule
	\end{tabularx}
\end{table*}

\begin{table*}[!h]
	\caption{Characteristics of standard and generalized ACC designs}  \label{Tab:Comparison}
	\begin{tabularx}{1\textwidth}{lXX}
		\toprule
		Characteristic & Standard Design & Generalized Design\\
		\midrule
		State distribution
		& All states follow the same type of distribution that may differ in scale parameters but not shape or location parameters.
		& States can follow different distributions and can differ in the scale, shape, or location parameters.\\
		Angles of ACLs
		& Each ACL has the same angle of inclination for all state transitions.
		& Each ACL has different angles for each state transition, depending on its distribution.\\
		Geometry of ACLs
		& Each ACL is a continuous straight line.
		& Each ACL is made up of several connected linear segments, one for each state transition.\\
		\bottomrule
	\end{tabularx}
\end{table*}

\section{Conclusions} \label{Sec:Conclusions}
This paper presents a new design of control charts to monitor the reliability of MSSs. This design is termed here as the Angular Control Chart. Two versions of the ACC design have been explored, the standard version and the generalized version. Furthermore, an open-source software named \emph{AC Charts Software} was developed to automate the implementation of the proposed ACC design. The ACC design replaces $t$- and $t_r$-charts of the binary-state system when the system becomes multi-state in failure. The ACC design has been examined with state transitions having different continuous probability distributions. However, if the ACC chart is used to monitor the cumulative TTF every $r$ occurrences of each state (Example II), further investigation is still needed to study the effect of the value of $r$ on the sensitivity of the ACC chart. (The latter use is restricted to a state transition distribution with a limiting sum distribution.) The practicality and limitations of the ACC design have been also discussed. The salient points of the practicality of the ACC design are its capability to monitor the state transitions individually and the system as a whole; and that distributions of the state transitions aren't restricted except to be non-negative and continuous. Two obstacles are still under study—clarifying the chronological order of the failure events and potential coincidences of two or more observation points. The ACC design opens a new direction in monitoring the reliability of MSSs.

\bibliographystyle{spbasic}
\bibliography{references}

\begin{thebibliography}{34}
\providecommand{\natexlab}[1]{#1}
\providecommand{\url}[1]{{#1}}
\providecommand{\urlprefix}{URL }
\expandafter\ifx\csname urlstyle\endcsname\relax
  \providecommand{\doi}[1]{DOI~\discretionary{}{}{}#1}\else
  \providecommand{\doi}{DOI~\discretionary{}{}{}\begingroup
  \urlstyle{rm}\Url}\fi
\providecommand{\eprint}[2][]{\url{#2}}

\bibitem[{Ali et~al.(2016)Ali, Pievatolo, and G{\"o}b}]{Ali.2016}
Ali S, Pievatolo A, G{\"o}b R (2016) An overview of control charts for
  high-quality processes. Quality and Reliability Engineering International
  32(7):2171--2189, \doi{10.1002/qre.1957}

\bibitem[{Alsyouf et~al.(2015)Alsyouf, Shamsuzzaman, Al-Taha, and
  Abdelrahman}]{Alsyouf.32015}
Alsyouf I, Shamsuzzaman M, Al-Taha M, Abdelrahman G (2015) Design of control
  chart for monitoring time-between-failures of a repairable system - a case
  study. In: 2015 International Conference on Industrial Engineering and
  Operations Management (IEOM), IEEE, pp 1--6, \doi{10.1109/IEOM.2015.7093794}

\bibitem[{Batson et~al.(2006)Batson, Jeong, Fonseca, and Ray}]{Batson.2006}
Batson RG, Jeong Y, Fonseca DJ, Ray PS (2006) Control charts for monitoring
  field failure data. Quality and Reliability Engineering International
  22(7):733--755, \doi{10.1002/qre.725}

\bibitem[{Besterfield(2013)}]{Besterfield.2013}
Besterfield DH (2013) Quality Improvement, 9th edn. {Prentice Hall}, New
  Jersey, USA

\bibitem[{Bury(1999)}]{Bury.1999}
Bury K (1999) Statistical Distributions in Engineering. {Cambridge University
  Press}, Cambridge, UK

\bibitem[{Calvin(1983)}]{Calvin.1983}
Calvin T (1983) Quality control techniques for {\textquotedbl}zero
  defects{\textquotedbl}. IEEE Transactions on Components, Hybrids, and
  Manufacturing Technology 6(3):323--328, \doi{10.1109/TCHMT.1983.1136174}

\bibitem[{Chan et~al.(2000)Chan, Xie, and Goh}]{Chan.2000}
Chan LY, Xie M, Goh TN (2000) Cumulative quantity control charts for monitoring
  production processes. International Journal of Production Research
  38(2):397--408, \doi{10.1080/002075400189482}

\bibitem[{Chou et~al.(1998)Chou, Polansky, and Mason}]{Chou.1998}
Chou YM, Polansky AM, Mason RL (1998) Transforming non-normal data to normality
  in statistical process control. Journal of Quality Technology 30(2):133--141,
  \doi{10.1080/00224065.1998.11979832}

\bibitem[{Fang et~al.(2016)Fang, Khoo, Teh, and Xie}]{Fang.2016}
Fang YY, Khoo MBC, Teh SY, Xie M (2016) Monitoring of time-between-events with
  a generalized group runs control chart. Quality and Reliability Engineering
  International 32(3):767--781, \doi{10.1002/qre.1789}

\bibitem[{Farouk et~al.(2011)Farouk, Soltan, Moughith, and Fikry}]{Farouk.2011}
Farouk M, Soltan HA, Moughith W, Fikry A (2011) Control charts for monitoring
  degraded reliability. In: Proceedings of the 5th International Conference on
  Mechanical Engineering Advanced Technology for Industrial Production
  (MEATIP5), {Assiut University}

\bibitem[{Forbes et~al.(2011)Forbes, Evans, Hastings, and
  Peacock}]{Forbes.2011}
Forbes C, Evans M, Hastings N, Peacock B (2011) Statistical Distributions, 4th
  edn. Wiley, Oxford, UK

\bibitem[{Genada et~al.(2015)Genada, Hussein, and Abdel-Shafi}]{Genada.2015}
Genada K, Hussein M, Abdel-Shafi A (2015) Control charts for monitoring the
  reliability of multi-state systems. Mansoura Engineering Journal
  40(4):50--57, \doi{10.21608/bfemu.2020.102395}

\bibitem[{Janada(2022)}]{Janada.2022}
Janada K (2022) {AC Charts Software v1.1}. Zenodo, \doi{10.5281/zenodo.6512066}

\bibitem[{Lisnianski et~al.(2010)Lisnianski, Frenkel, and
  Ding}]{Lisnianski.2010}
Lisnianski A, Frenkel I, Ding Y (2010) Multi-State System Reliability Analysis
  and Optimization for Engineers and Industrial Managers. {Springer London},
  London, UK, \doi{10.1007/978-1-84996-320-6}

\bibitem[{Liu et~al.(2006)Liu, Xie, Goh, and Sharma}]{Liu.2006}
Liu JY, Xie M, Goh TN, Sharma PR (2006) A comparative study of exponential time
  between events charts. Quality Technology {\&} Quantitative Management
  3(3):347--359, \doi{10.1080/16843703.2006.11673120}

\bibitem[{Natvig(2010)}]{Natvig.2010}
Natvig B (2010) Multistate system reliability. In: Cochran JJ, Cox LA,
  Keskinocak P, Kharoufeh JP, Smith JC (eds) Wiley Encyclopedia of Operations
  Research and Management Science, {John Wiley {\&} Sons, Inc}, New Jersey,
  USA, \doi{10.1002/9780470400531.eorms0553}

\bibitem[{Nelson(1994)}]{Nelson.1994}
Nelson LS (1994) A control chart for parts-per-million nonconforming items.
  Journal of Quality Technology 26(3):239--240,
  \doi{10.1080/00224065.1994.11979529}

\bibitem[{O'Connor et~al.(2016)O'Connor, Modarres, and Mosleh}]{OConnor.2011}
O'Connor AN, Modarres M, Mosleh A (2016) Probability Distributions Used in
  Reliability Engineering. {University of Maryland Center for Risk and
  Reliability (RiAC)}, Maryland, USA

\bibitem[{Rasay and Arshad(2020)}]{Rasay.2020}
Rasay H, Arshad H (2020) Designing variable control charts under failure
  censoring reliability tests with replacement. Transactions of the Institute
  of Measurement and Control 42(15):3002--3011, \doi{10.1177/0142331220938206}

\bibitem[{Sabri-Laghaie et~al.(2022)Sabri-Laghaie, Fathi, Zio, and
  Mazhar}]{SabriLaghaie.2022}
Sabri-Laghaie K, Fathi M, Zio E, Mazhar M (2022) A novel reliability monitoring
  scheme based on the monitoring of manufacturing quality error rates.
  Reliability Engineering {\&} System Safety 217:108065,
  \doi{10.1016/j.ress.2021.108065}

\bibitem[{Sharma et~al.(2006)Sharma, Xie, and Goh}]{Sharma.2006}
Sharma PR, Xie M, Goh TN (2006) Monitoring inter-arrival times with statistical
  control charts. In: Pham H (ed) Reliability Modeling, Analysis and
  Optimization, Series on Quality, Reliability and Engineering Statistics,
  vol~9, {World Scientific}, pp 43--66, \doi{10.1142/9789812707147_0003}

\bibitem[{Shore(2001)}]{Shore.2001}
Shore H (2001) Process control for non-normal populations based on an inverse
  normalizing transformation. In: Lenz HJ, Wilrich PT (eds) Frontiers in
  Statistical Quality Control 6, Springer-Verlag, Berlin, Germany, pp 194--206,
  \doi{10.1007/978-3-642-57590-7_12}

\bibitem[{Soltan(2019)}]{Soltan.2019}
Soltan H (2019) Advances in control charts for reliability. In: Industrial {\&}
  Systems Engineering Conference (ISEC), IEEE, pp 1--4,
  \doi{10.1109/IASEC.2019.8686608}

\bibitem[{S{\"u}r{\"u}c{\"u} and Sazak(2009)}]{Surucu.2009}
S{\"u}r{\"u}c{\"u} B, Sazak HS (2009) Monitoring reliability for a three
  parameter {Weibull} distribution. Reliability Engineering {\&} System Safety
  94(2):503--508, \doi{10.1016/j.ress.2008.06.001}

\bibitem[{Wu et~al.(2020)Wu, Younas, Abbas, Ali, and Khan}]{Wu.2020}
Wu Y, Younas S, Abbas K, Ali A, Khan SA (2020) Monitoring reliability for
  three-parameter {Frechet} distribution using control charts. IEEE Access
  8:71245--71253, \doi{10.1109/ACCESS.2020.2987422}

\bibitem[{Wu et~al.(2009{\natexlab{a}})Wu, Jiao, and He}]{Wu.2009}
Wu Z, Jiao J, He Z (2009{\natexlab{a}}) A control scheme for monitoring the
  frequency and magnitude of an event. International Journal of Production
  Research 47(11):2887--2902, \doi{10.1080/00207540701689743}

\bibitem[{Wu et~al.(2009{\natexlab{b}})Wu, Jiao, and He}]{Wu.2009b}
Wu Z, Jiao J, He Z (2009{\natexlab{b}}) A single control chart for monitoring
  the frequency and magnitude of an event. International Journal of Production
  Economics 119(1):24--33, \doi{10.1016/j.ijpe.2009.01.004}

\bibitem[{Xie et~al.(2002{\natexlab{a}})Xie, Goh, and
  Kuralmani}]{Xie.2002.Book}
Xie M, Goh TN, Kuralmani V (2002{\natexlab{a}}) Statistical Models and Control
  Charts for High-Quality Processes. {Springer US}, Boston, USA,
  \doi{10.1007/978-1-4615-1015-4}

\bibitem[{Xie et~al.(2002{\natexlab{b}})Xie, Goh, and
  Ranjan}]{Xie.2002.Article}
Xie M, Goh TN, Ranjan P (2002{\natexlab{b}}) Some effective control chart
  procedures for reliability monitoring. Reliability Engineering {\&} System
  Safety 77(2):143--150, \doi{10.1016/S0951-8320(02)00041-8}

\bibitem[{Xie et~al.(2011)Xie, Goh, and Deng}]{Xie.092011}
Xie M, Goh TN, Deng PP (2011) A circle chart for periodic measurements. In:
  2011 IEEE International Conference on Quality and Reliability (ICQR), IEEE,
  pp 595--599, \doi{10.1109/ICQR.2011.6031609}

\bibitem[{Xie et~al.(2012)Xie, Goh, and Deng}]{Xie.2012}
Xie M, Goh TN, Deng PP (2012) Circle chart for monitoring of periodic
  measurements. Quality and Reliability Engineering International
  28(8):943--948, \doi{10.1002/qre.1285}

\bibitem[{Yang and Xie(2000)}]{Yang.2000}
Yang Z, Xie M (2000) Process monitoring of exponentially distributed
  characteristics through an optimal normalizing transformation. Journal of
  Applied Statistics 27(8):1051--1063, \doi{10.1080/02664760050173373}

\bibitem[{Yingkui and Jing(2012)}]{Yingkui.2012}
Yingkui G, Jing L (2012) Multi-state system reliability: a new and systematic
  review. Procedia Engineering 29:531--536, \doi{10.1016/j.proeng.2011.12.756}

\bibitem[{Zhang et~al.(2011)Zhang, Xie, Goh, and Shamsuzzaman}]{Zhang.2011}
Zhang HY, Xie M, Goh TN, Shamsuzzaman M (2011) Economic design of time between
  events control chart system. Computers {\&} Industrial Engineering
  60(4):485--492, \doi{10.1016/j.cie.2010.11.008}

\end{thebibliography}
\end{document}